\documentclass[aps,prd,showpacs,preprintnumbers,groupedaddress,nofootinbib]{revtex4}
\usepackage{amsmath}
\usepackage{graphicx}
\usepackage{bm}
\usepackage{color,colordvi}
\usepackage{feynarts}
\begin{document}

\def\cM{{\cal M}} 
\def\cO{{\cal O}}
\def\cK{{\cal K}}
\def\cS{{\cal S}}
\newcommand{\mh}{m_h}
\newcommand{\mw}{m_W}
\newcommand{\mz}{m_Z}
\newcommand{\mt}{m_t}
\newcommand{\mb}{m_b}
\def\lsim{\mathrel{\raise.3ex\hbox{$<$\kern-.75em\lower1ex\hbox{$\sim$}}}}
\def\gsim{\mathrel{\raise.3ex\hbox{$>$\kern-.75em\lower1ex\hbox{$\sim$}}}}
\def\ga{\mathrel{\raise.3ex\hbox{$>$\kern-.75em\lower1ex\hbox{$\sim$}}}}
\def\la{\mathrel{\raise.3ex\hbox{$<$\kern-.75em\lower1ex\hbox{$\sim$}}}}
\newcommand{\gznul}{g_Z^{\nu{L}}}
\newcommand{\gzel}{g_Z^{eL}}
\newcommand{\gzer}{g_Z^{eR}}
\newcommand{\gzdl}{g_Z^{dL}}
\newcommand{\gzdr}{g_Z^{dR}}
\newcommand{\gzul}{g_Z^{uL}}
\newcommand{\gzur}{g_Z^{uR}}
\newcommand{\gpnul}{g_{\gamma}^{\nu{L}}}
\newcommand{\gpel}{g_{\gamma}^{eL}}
\newcommand{\gper}{g_{\gamma}^{eR}}
\newcommand{\gpdl}{g_{\gamma}^{dL}}
\newcommand{\gpdr}{g_{\gamma}^{dR}}
\newcommand{\gpul}{g_{\gamma}^{uL}}
\newcommand{\gpur}{g_{\gamma}^{uR}}
\newcommand{\yl}{y_{Htb}^L}
\newcommand{\yr}{y_{Htb}^R}
\newcommand{\ygl}{y_{Gtb}^L}
\newcommand{\ygr}{y_{Gtb}^R}
\newcommand{\grvtt}{g^{R}_{Vtt}}
\newcommand{\grvbb}{g^{R}_{Vbb}}
\newcommand{\glvtt}{g^{L}_{Vtt}}
\newcommand{\glvbb}{g^{L}_{Vbb}}
\newcommand{\glwtb}{g^{L}_{W}}
\newcommand{\glwtbc}{g^{L*}_{W}}
\newcommand{\glhtb}{g^{L}_{H}}
\newcommand{\grhtb}{g^{R}_{H}}

\newcommand{\aone}{{\cal{A}}_1}
\newcommand{\atwo}{{\cal{A}}_2}
\newcommand{\athree}{{\cal{A}}_3}
\newcommand{\afour}{{\cal{A}}_4}
\newcommand{\afive}{{\cal{A}}_5}
\newcommand{\asix}{{\cal{A}}_6}
\newcommand{\aseven}{{\cal{A}}_7}
\newcommand{\aeight}{{\cal{A}}_8}

\newcommand{\cw}{c_{W}}
\newcommand{\sw}{s_{W}}
\newcommand{\cba}{c_{\beta\alpha}}
\newcommand{\sba}{s_{\beta\alpha}}
\newcommand{\sinb}{\sin\beta}
\newcommand{\cosb}{\cos\beta}
\newcommand{\sat}{\sin\theta_f}
\newcommand{\cat}{\cos\theta_f}
\newcommand{\sab}{\sin\theta_f}
\newcommand{\cab}{\cos\theta_f}

\newcommand{\ghag}{g_{H^{-}AG^{+}}}
\newcommand{\ghhh}{g_{H^-h^{0}H^+}}
\newcommand{\ghHh}{g_{H^-H^{0}H^+}}
\newcommand{\ghhg}{g_{H^-h^{0}G^+}}
\newcommand{\ghHg}{g_{H^-H^{0}G^+}}
\newcommand{\gghg}{g_{G^-h^{0}G^+}}
\newcommand{\ggHg}{g_{G^-H^{0}G^+}}
\newcommand{\ghghh}{g_{H^-G^{+}h^0h^0}}
\newcommand{\ghgHH}{g_{H^-G^{+}H^0H^0}}
\newcommand{\ghgaa}{g_{H^-G^{+}A^0A^0}}
\newcommand{\ghghpmhpm}{g_{H^-G^{+}H^+H^-}}
\newcommand{\ghggg}{g_{H^-G^{+}G^+G^-}}

\newcommand{\gw}{g_W}
\newcommand{\gpl}{g_{\gamma}^L}
\newcommand{\gzl}{g_{Z}^L}
\newcommand{\gpr}{g_{\gamma}^R}
\newcommand{\gzr}{g_{Z}^R}
\newcommand{\gvl}{g_{V}^L}
\newcommand{\ghl}{g_{H}^L}
\newcommand{\ggl}{g_{G}^L}
\newcommand{\gwl}{g_{W}^L}
\newcommand{\gvr}{g_{V}^R}
\newcommand{\ghr}{g_{H}^R}
\newcommand{\gwr}{g_{W}^R}
\newcommand{\ggr}{g_{G}^R}

\newcommand{\mi}{m_{\chi_i}}
\newcommand{\mj}{m_{\chi_j}}
\newcommand{\mk}{m_{\chi_k}}
\newcommand{\mitwo}{m_{\chi_i}^2}
\newcommand{\mjtwo}{m_{\chi_j}^2}
\newcommand{\mktwo}{m_{\chi_k}^2}
\newcommand{\msltwo}{m_{\tilde{l}}^2}
\newcommand{\mseltwo}{m_{\tilde{e}_L}^2}
\newcommand{\msnutwo}{m_{\tilde{\nu}}^2}

\newcommand{\slepton}{{\tilde{l}}}
\newcommand{\snu}{{\tilde{\nu}}}
\newcommand{\sel}{{\tilde{e}}_L}
\newcommand{\ser}{{\tilde{e}}_R}
\newcommand{\sul}{{\tilde{u}}_L}
\newcommand{\sdl}{{\tilde{d}}_L}
\newcommand{\stone}{{\tilde{t}}_{1}}
\newcommand{\sttwo}{{\tilde{t}}_{2}}
\newcommand{\sbone}{{\tilde{b}}_{1}}
\newcommand{\sbtwo}{{\tilde{b}}_{2}}
\newcommand{\sbl}{{\tilde{b}}_L}
\newcommand{\sbr}{{\tilde{b}}_R}
\newcommand{\stl}{{\tilde{t}}_L}
\newcommand{\str}{{\tilde{t}}_R}
\newcommand{\ghenu}{g_{H\sel\snu}}
\newcommand{\ghdu}{g_{H\sdl\sul}}
\newcommand{\ggenu}{g_{G\sel\snu}}
\newcommand{\ggdu}{g_{G\sdl\sul}}
\newcommand{\ghbtone}{g_{H\sbl\stone}}
\newcommand{\ghbttwo}{g_{H\sbl\sttwo}}
\newcommand{\ghbtoner}{g_{H\sbr\stone}}
\newcommand{\ghbttwor}{g_{H\sbr\sttwo}}
\newcommand{\ggbtone}{g_{G\sbl\stone}}
\newcommand{\ggbttwo}{g_{G\sbl\sttwo}}
\newcommand{\ggbtoner}{g_{G\sbr\stone}}
\newcommand{\ggbttwor}{g_{G\sbr\sttwo}}
\newcommand{\ghbonetone}{g_{H\sbone\stone}}
\newcommand{\ghbonettwo}{g_{H\sbone\sttwo}}
\newcommand{\ghbtwotone}{g_{H\sbtwo\stone}}
\newcommand{\ghbtwottwo}{g_{H\sbtwo\sttwo}}
\newcommand{\ggbonetone}{g_{G\sbone\stone}}
\newcommand{\ggbonettwo}{g_{G\sbone\sttwo}}
\newcommand{\ggbtwotone}{g_{G\sbtwo\stone}}
\newcommand{\ggbtwottwo}{g_{G\sbtwo\sttwo}}

\newcommand{\gchinur}{g_{{\chi}\nu\tilde{l}}^R}
\newcommand{\gchiel}{g_{{\chi}e\tilde{l}}^L}
\newcommand{\gchiie}{g_{{\chi_i}e\tilde{l}}}
\newcommand{\gchije}{g_{{\chi_j}e\tilde{l}}}
\newcommand{\gchike}{g_{{\chi_k}e\tilde{l}}}
\newcommand{\gchiiel}{g_{{\chi_i}e\tilde{l}}^L}
\newcommand{\gchiier}{g_{{\chi_i}e\tilde{l}}^R}
\newcommand{\gwsesnu}{g_{W\sel\snu}}
\newcommand{\gchijel}{g_{{\chi_j}e\tilde{l}}^L}
\newcommand{\gchijer}{g_{{\chi_j}e\tilde{l}}^R}
\newcommand{\gchijnur}{g_{{\chi_j}\nu\tilde{l}}^R}
\newcommand{\gchiiesel}{g_{{\chi_i}e\tilde{e}_L}^L}
\newcommand{\gchiinusnur}{g_{{\chi_i}\nu\tilde{\nu}}^R}
\newcommand{\gchiieser}{g_{{\chi_i}e\tilde{e}_L}^R}
\newcommand{\gchiiesnul}{g_{{\chi_i}e\tilde{\nu}}^L}
\newcommand{\gchiiesnur}{g_{{\chi_i}e\tilde{\nu}}^R}
\newcommand{\gchijeser}{g_{{\chi_j}e\tilde{e}_L}^R}
\newcommand{\gchijesel}{g_{{\chi_j}e\tilde{e}_L}^L}
\newcommand{\tlf}{\tilde{f}}
\newcommand{\mhp}{m_{H^{\pm}}}
\newcommand{\non}{\nonumber}
\noindent
$\ $\\
$\ $\\
$\ $\\
\preprint{LPHEA-06-03}

\title{Charged Higgs Bosons decays $H^\pm \to W^\pm (\gamma ,Z)$ revisited }

\author{Abdesslam Arhrib$^{1,2}$}
\email[]{aarhrib@ictp.it}
\affiliation{$^1$D\'epartement de Math\'ematiques, Facult\'e des Sciences et 
Techniques
B.P 416 Tanger, Morocco.} 
\author{Rachid Benbrik$^2$}
\email[]{r.benbrik@ucam.ac.ma}
\affiliation{$^2$LPHEA, D\'epartement de Physique,
             Facult\'e des Sciences-Semlalia,
             B.P. 2390 Marrakech, Morocco.}
\author{Mohamed Chabab$^2$}
\email[]{mchabab@ucam.ac.ma}
\affiliation{$^2$LPHEA, D\'epartement de Physique,
             Facult\'e des Sciences-Semlalia,
             B.P. 2390 Marrakech, Morocco.}
\noindent
\vspace{1.cm}

\begin{abstract}
We study the complete one loop contribution to
$H^\pm\to W^\pm V$, $V= Z, \gamma$, both in the Minimal 
Supersymmetric Standard Model (MSSM) and in the 
Two Higgs Doublet Model (2HDM). We evaluate the 
MSSM contributions and compare them with the 2HDM ones
taking into account $b\to s\gamma$ constraint, vacuum stability and 
unitarity constraints in the case of 2HDM, as well as experimental 
constraints on the MSSM and 2HDM parameters. 
In the MSSM, we found that in the intermediate range of 
$\tan\beta \la 10$ and for large $A_t$, 
the branching ratio of $H^\pm \to W^{\pm} Z$ can be of the order 
$10^{-3}$ while the branching ratio of $H^\pm \to W^{\pm} \gamma$ 
is of the order $10^{-5}$.
We also study the effects of the CP violating phases of Soft SUSY
parameters and found that they can modify the branching 
ratio by about one order of magnitude.
However, in the 2HDM where the Higgs sector is less 
constrained as compared to the MSSM higgs sector, 
one can reach branching ratio of the order $10^{-2}$ for both modes. 
\end{abstract}

\maketitle

\newpage
\section{Introduction}
\label{sec:intro}
The Standard Model (SM) of electroweak interactions \cite{Wein} 
is very successful in  explaining all experimental data available
till now. The cornerstone of the SM, the electroweak symmetry breaking 
mechanism, still has to be established and the Higgs boson has 
to be discovered.
The main goals of future colliders such as LHC and ILC is to 
study the scalar sector of the SM.
Moreover, the problematic scalar sector of the SM can be 
enlarged and some simple extensions such as the
 Minimal Supersymmetric Standard Model (MSSM) and the 
Two Higgs Doublet Model (2HDM) \cite{HHG,abdel2} are intensively
studied. Both in the 2HDM and MSSM the electroweak symmetry breaking is 
generated by 2 Higgs doublets fields $\Phi_1$ and $\Phi_2$. 
After electroweak symmetry
breaking we are left with 5 physical Higgs particles 
(2 charged Higgs $H^\pm$, 2 CP-even $H^0$, $h^0$
and one CP--odd $A^0$). The charged Higgs $H^\pm$, 
because of its electrical charge,  
is noticeably different from the other SM or 2HDM/MSSM Higgs
particles, its discovery would be a clear 
evidence of physics beyond the SM.

The charged Higgs can be copiously produced both at hadrons and 
 $e^+e^-$ colliders. In hadronic machines, the charged Higgs bosons 
can be produced in many channels:
$i$) the production of $t\bar{t}$ pairs may offer a
source of charged Higgs production. If kinematicaly allowed
$m_{H^\pm}\la m_t$, 
the top quark can decay to $H^+\bar{b}$, 
competing with the SM decay $t\to W^+b$. 
This mechanism can provide a larger production rate of charged Higgs
and offers a much cleaner signature than that of direct production.\\
$ii$) single charged Higgs production via $gb \to tH^-$,
$gg\to t\bar{b}H^-$, $qb\to q' bH^-$ \cite{single}.
$iii$) single charged Higgs production in association with 
$W^\pm$ gauge boson via
$ gg \to W^\pm H^\mp $ or $b\bar{b} \to  W^\pm H^\mp $ \cite{wh}
and also single charged Higgs production in association with 
$A^0$ boson via $qq, gg \to A^0 H^\mp$ \cite{cpyuan}.
$iv$) $H^\pm $ pair production through $q\bar{q}$ 
annihilation \cite{hadronic} or gluon fusion.\\
At $e^+e^-$ colliders, the simplest way to get a charged Higgs is 
through $H^\pm$  pair production. 
Such studies have been already undertaken at tree-level \cite{komamiya}
and one-loop orders \cite{ACM} and shown that
$e^+e^-$ machines will offer a clean environment and in that sense 
a higher mass reach.  We mention also that charged Higgs bosons 
pair production through laser back--scattered $\gamma\gamma$ 
collisions has been studied in the literature \cite{pp} 
and found to be prominent to discover the charged Higgs boson.

Experimentally, the null--searches from L3  
collaborations at LEP-II derive the lower limit of about 
$m_{H^{\pm}}\ga 80$ GeV  \cite{LEP1}, a limit
which applies to all models (2HDM or MSSM) in which 
BR($H^{\pm}\to \tau\nu_{\tau}$)+
BR($H^{\pm}\to cs$)=1. DELPHI has also carried out 
search for $H^{\pm}\to A^0W^\pm$ \footnote{Note that in the 2HDM it 
may be possible that the decay channel $H^\pm\to W^\pm A^0$ is open 
and even dominate over $\tau\nu$ mode for $m_{H\pm}\la m_t$ 
\cite{borzumati,andrew,aan}. } 
topologies in the context of 2HDM
type I and derive the lower limit of about 
$m_{H^{\pm}}\ga 76$ GeV \cite{LEP2}.
Recently, CDF Run II excluded a charged Higgs mass in the range 
$80< m_{H^\pm}< 160$ GeV \cite{tev}. This limit can apply for the 
MSSM with low $\tan\beta$. If the charged Higgs decay 
exclusively to $\bar{\tau}\nu$ the BR$(t\to H^{+}b$) is constrained to
be less than $0.4$ at 95\%C.L. On the other hand if no assumption is
made on charged Higgs decay, the BR$(t\to H^{+}b$) is constrained to
be less than $0.91$ at 95\%C.L.

At the LHC, the detection of light charged Higgs boson with
$m_{H^\pm}\la m_t$ is straightforward from top production 
followed by  the decay $t\to
bH^+$\footnote{Note that at Tevatron run II, the charged Higgs is also
searched in top decay \cite{tev}.}.
Such light charged Higgs ($m_{H^\pm}\la m_t$) can be detected also 
for any $\tan\beta$ in
the $\tau\nu$ decay  which is indeed the dominant decay mode 
\cite{chaud}.
However, for heavy charged Higgs masses $m_{H^\pm}\ga m_t$
which decay predominantly to $t\bar{b}$, 
the search is rather difficult due to large irreducible and reducible
backgrounds associated with $H^+\to t\bar{b}$ decay.
However, it has been demonstrated in \cite{Htb} that the 
 $H^+\to t\bar{b}$ signature can lead to a visible signal at LHC
provided that the charged Higgs mass below 600 GeV and 
$\tan\beta$ is either below $\la 1.5$ or above $\ga 40$.
Ref.~\cite{odagiri}, proposed $H^\pm\to \tau\nu$ as an alternative
decay mode to detect a heavy charged Higgs, even if such decay is 
suppressed for heavy charged Higgs it has the advantage being more
clean than  $H^+\to t\bar{b}$. \\
An other alternative discovery channel for heavy charged Higgs is its 
decay to charged gauge boson and lightest CP-even Higgs: $H^\pm \to W^\pm h^0$,
followed by the dominant decay of $h^0$ to $b\bar{b}$ \cite{moretti}. 
Since the branching ratio of $H^\pm \to W^\pm h^0$ is suppressed for
High $\tan\beta$, this channel could lead to charged Higgs discovery
only for low $\tan\beta$  where the branching ratio of $H^\pm \to
W^\pm h^0$ is sizeable.

Both in 2HDM as well as in MSSM, at tree level, 
the coupling $H^\pm \to W^\pm \gamma$
is absent because of electromagnetic gauge invariance $U(1)_{\rm em}$.
While the absence of $H^\pm \to W^\pm Z $ is due to the 
isospin symmetry of the kinetic Lagrangian of the Higgs fields 
\cite{iso}.
Therefore, decays modes like $H^\pm \to W^\pm \gamma $, 
$H^\pm \to W^\pm Z $ are mediated at one loop level and then are
expected to be loop suppressed \cite{pom,micapey,ray,kanemur,toscano}. 
We emphasize here that it is possible to 
construct models with an even larger scalar sector than 
2 Higgs doublets, one of the most popular being the Higgs Triplet
Model (HTM) \cite{triplet}. 
A noteworthy difference between 2HDM and HTM is that 
the HTM contains a tree level $ZW^\pm H^\mp$ coupling.\\
Motivated by the fact that there is no 
detailed study about $H^\pm \to W^\pm V$, $V=Z,\gamma$, 
in the the framework of MSSM in the literature 
which take into account $b\to s \gamma$ and other electroweak
and experimental constraints. We would like to reconsider 
and update the existing works 
\cite{pom,micapey,ray,kanemur,toscano} on the charged Higgs boson decays
into a pair of gauge boson: $H^\pm \to W^\pm \gamma , W^\pm
Z$ both in 2HDM and MSSM with and without CP violating phases.
Although these decays are rare processes, 
loop or/and threshold effects can give a substantial effect.
Moreover, once worked out, any experimental deviation from the results 
within such a model should bring some fruitful information on the new 
physics and allow to distinguish between models. 
We would like to mention also that, those channels have
a very clear signature and might emerge easily at future colliders.
For instance, if $H^\pm \to W^\pm Z $ is enhanced enough, this decay 
may lead to three leptons final state if both W and Z decay leptonically
and that would be the corresponding golden mode for charged Higgs boson.
 
Charged Higgs decays: $H^\pm \to W^\pm \gamma , W^\pm Z$, have
received much more attention in the literature.
$H^\pm \to W^\pm Z$ has been studied first in the MSSM in \cite{pom}.
Ref.~\cite{micapey} has considered both 
$H^\pm \to W^\pm \gamma$  and $H^\pm \to W^\pm Z$ in the MSSM
and show that the rate of $H^\pm \to W^\pm \gamma$ is very small
while the rate of $H^\pm \to W^\pm Z$ can be enhanced by 
heavy fermions particles in the loops. The fourth generation
contribution was given as an example. 
Although the squarks contribution has been considered in 
Ref.~\cite{micapey}, Left-Right mixing which could give substantial
enhancement has been neglected.
$H^\pm \to W^\pm \gamma$ was also studied in 
\cite{ray} within the MSSM.
Later on, Ref.~\cite{kanemur}
studied the possibility of enhancing $H^\pm \to W^\pm Z$ by the
non-decoupling effect of the heavy Higgs bosons 
in the context of 2HDM, substantial enhancement was found 
\cite{kanemur}. Recently, $H^\pm \to W^\pm \gamma $
was also studied in 2HDM type II \cite{toscano}.
All the above studies has been carried out either in unitary gauge
\cite{pom,micapey} or in the nonlinear 
$R_{\xi}$-gauge \cite{toscano}. The analysis of \cite{ray} and 
\cite{kanemur} have been performed in `tHooft-Feynman gauge without
any renormalization scheme. It has been checked in \cite{ray,kanemur}
that the sum of all Feynman diagrams: vertex, tadpoles and
vector boson--scalar mixing turns out to be Ultra-Violet finite. \\
In the present study, we will still use `tHooft-Feynman gauge to do
the computation. However, the amplitudes of $H^\pm \to W^\pm \gamma $
and $H^\pm \to W^\pm Z $ are absent at the tree level,
complications like tadpoles contributions and 
vector boson--scalar mixing require a careful treatment of 
renormalization. We adopt  hereafter the on-shell renormalization 
scheme developed in \cite{achm}.\\
The paper is organized as follows. In section II, we 
describe our calculations and the one-loop renormalization scheme
we will use for $H^\pm\to W^\pm Z$ and 
$H^\pm\to W^\pm \gamma$. In Section III, we present our numerical
results and discussions, and section VI contains our conclusions.

\section{Charged Higgs decay: $H^\pm\to W^\pm V$}

As we have seen in the previous section,
both in 2HDM as well as in MSSM, at tree level, 
the coupling $H^\pm \to W^\pm \gamma$ and $H^\pm \to W^\pm Z $
do not exist. They are  generated at one loop level and then are
expected to be loop suppressed \cite{pom,micapey,ray,kanemur,toscano}.
Hereafter, we will give the general structure of such 
one loop couplings and discuss the renormalization scheme introduced
to deal with tadpoles and vector boson scalar boson mixing.
\subsection{One loop amplitude $H^\pm\to W^\pm V$}
The amplitude ${\cal M}$ for a scalar decaying to two
gauges bosons $V_1$ and $V_2$ can be written as
\begin{eqnarray}
{\cal M}=\frac{g^3\epsilon_{V_1}^{\mu *}\epsilon_{V_2}^{\nu *}}
{16\pi^2 m_W}{\cal M}_{\mu\nu}
\end{eqnarray}
where $\epsilon_{V_i}$ are the polarization vectors of the $V_i$.\\
According to Lorenz invariance, the general structure of 
the one loop amplitude ${\cal M}_{\mu\nu}$ of 
$S\to V_1^{\mu} V_2^{\nu}$ decay,
if CP is conserved, is
\begin{eqnarray}
{\cal M}_{\mu\nu}(S\to W^\mu V^\nu) =
{\cal F}_1 g_{\mu\nu} + {\cal F}_2 p_{1\mu} p_{2\nu} + 
{\cal F}_3 
i\epsilon_{\mu\nu\rho\sigma}p_{1}^{\rho}p_{2}^{\sigma}\label{lor}
\end{eqnarray}
where $p_{1,2}$ are the momentum of $V_{1}$, $V_{2}$ vector bosons, 
${\cal F}_{1,2,3}$ are form factors, and 
$\epsilon_{\mu\nu\rho\sigma}$ is the totally antisymmetric tensor.
The form factor ${\cal F}_1$ has dimension 2 while the other are 
dimensionless.
The analytic expression for ${\cal F}_i$ are given in appendix B.

 For $H^\pm \to W^\pm \gamma$, electromagnetic gauge
invariance implies that 
${\cal F}_1=$$1/2(m_W^2-m_{H\pm}^2){\cal F}_2$ \cite{micapey}.
This means that only  ${\cal F}_2$ and ${\cal F}_3$ will 
contribute to the decay $H^\pm \to W^\pm \gamma$.
In case of $H^\pm \to W^\pm Z$, there is no such constraint on form
factors.\\
In terms of an effective Lagrangian analysis, from gauge invariance 
requirement we can write:
\begin{eqnarray}
{\cal L}_{eff} = g_1 H^\pm W_\mu^\mp V^\mu + g_2 H^\pm F_V^{\mu\nu} 
F_{W\mu\nu}
+ i g_3 \epsilon_{\mu\nu\rho\sigma} H^\pm F_V^{\mu\nu} F^{W\rho\sigma} 
+ h.c
\end{eqnarray}
the first operator $H^\pm W_\mu^\mp V^\mu$ is 
dimension three and the last two  operators
$H^\pm F_V^{\mu\nu} F_{W\mu\nu}$ and 
$\epsilon_{\mu\nu\rho\sigma}$ $H^\pm F_V^{\mu\nu} F^{W\rho\sigma}$
are dimension five. One conclude that $g_{2,3}$ (resp. $g_1$) 
must be of the form $g(R)/M$ (resp $M g(R)$) with $M$ a 
heavy scale in MSSM or in 2HDM, $g(R)$ a 
dimensionless function and $R$ is a ratio of some internal 
masses of the model under studies. 
Therefore, it is expected that in case of $H^\pm \to W^\pm Z$ decay, 
${\cal F}_1$ will grow quadratically with internal masses
while ${\cal F}_{2,3}$ will have only logarithmic dependence.
A contrario, for $H^\pm\to W^\pm \gamma$ decay, the
electromagnetic gauge invariance relates ${\cal F}_{1}$ and 
${\cal F}_{2}$ and then the amplitude of $H^\pm \to W^\pm \gamma$
will not grows quadratically with internal masses.
One expect that the decay $H^\pm \to W^\pm \gamma$ is less enhanced 
compared to $H^\pm \to W^\pm Z$.\\
After squaring the amplitude and summing over polarization vectors,
the decay widths as functions of the form factors ${\cal F}_1$, 
${\cal F}_2$ and ${\cal F}_3$ take the following form:
\begin{eqnarray}
\Gamma(H^{\pm}\to W^\pm Z)&=&\frac{g^6\,\lambda^{1/2}}
{2^{16}\pi^5\,m^6_W\,m^3_{H^\pm}}\Big[4(\lambda+12\,m^2_W\,m^2_Z)
|{\cal F}_1|^2+\lambda^2|{\cal F}_2|^2 + 
8\lambda\,m^2_W\,m^2_Z|{\cal F}_3|^2\\\non&+& 
4\lambda(m^2_{H^\pm}-m^2_W-m^2_Z)\Re e({\cal F}_1{\cal F}_2^*)\Big]\\
\Gamma(H^{\pm}\to W^\pm \gamma)&=&\frac{g^6\,m^3_{H^\pm}}
{2^{13}\pi^5\,m^2_W\,\cos^2\theta_W}
\Big[1-\frac{m^2_W}{m^2_{H^\pm}}\Big]^3\Big(|{\cal F}_1|^2+|{\cal F}_3|^2\Big)
\end{eqnarray}
where $\lambda = (m^2_{H^\pm} - m^2_W - m^2_Z)^2-4m^2_Wm^2_Z$
\begin{figure}[t!]
\begin{center}
\input{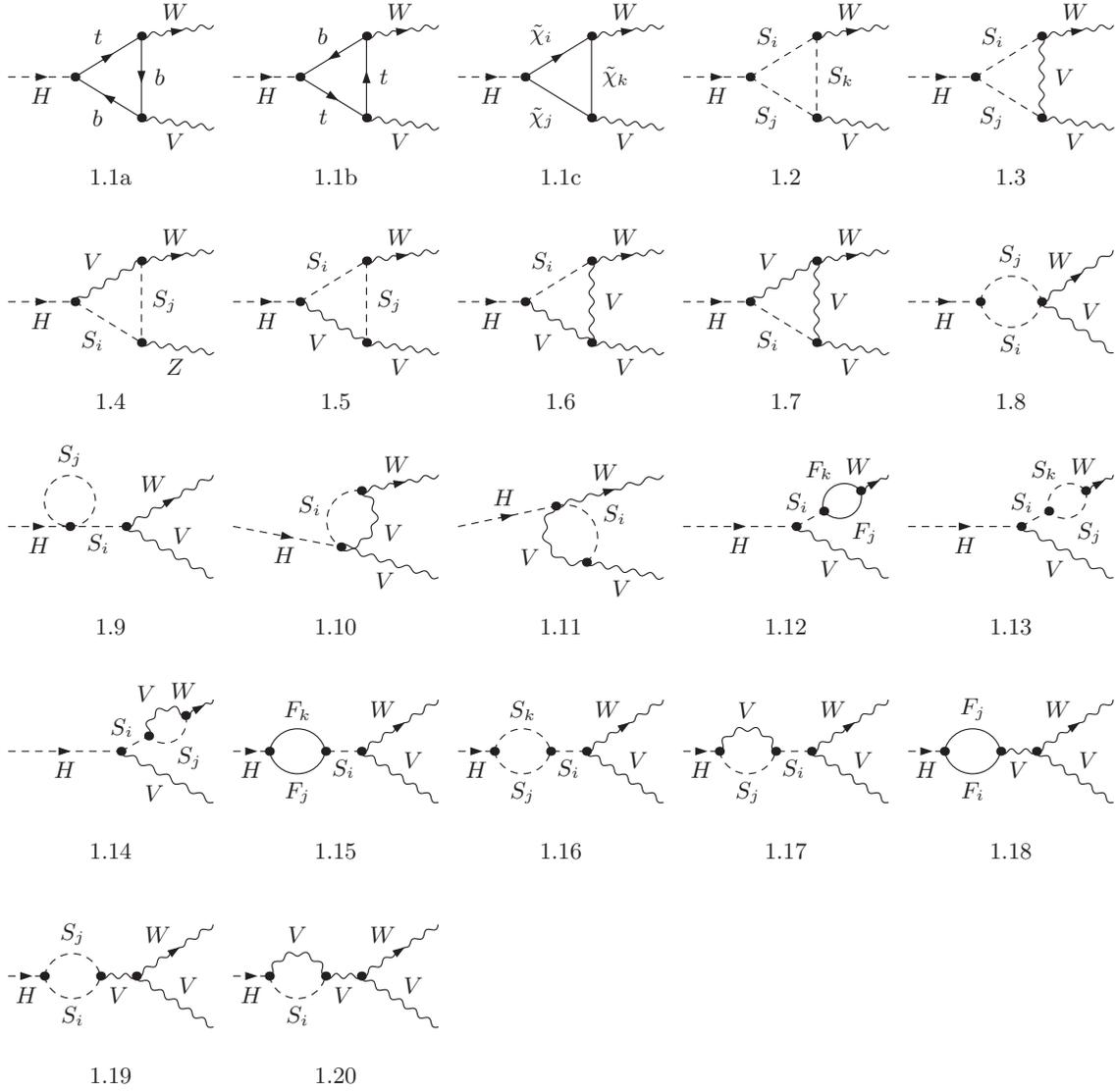}
\caption{Generic contributions to $H^\pm \to W^{\pm}V$}
\label{hwz}
\end{center}
\end{figure}
\begin{figure}[t!]
\begin{center}
\input{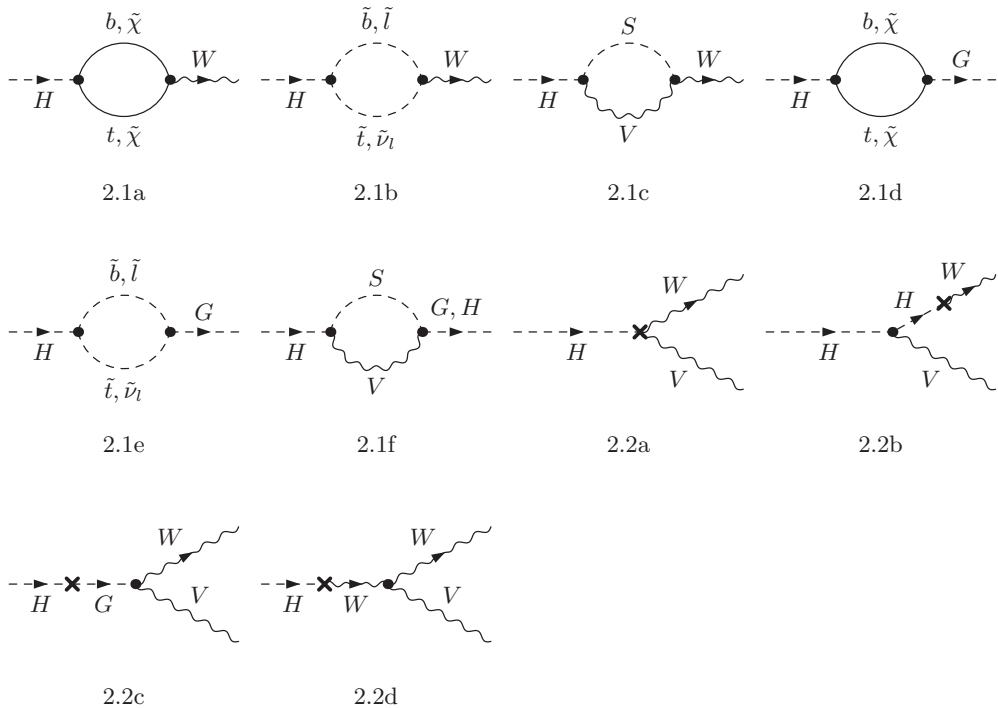}
\vspace{-6cm}
\caption{Generic contributions to $H^\pm \to W^{\pm}$ and
$H^\pm \to G^{\pm}$ mixing as well as counter-terms needed.}
\label{hwmix}
\end{center}
\end{figure}
\subsection{On-shell renormalization}
We have evaluated the one-loop induced process
$H^\pm \to  W^\pm V$ in the 'tHooft-Feynman gauge using
dimensional regularization. 
The  typical Feynman diagrams that contribute to  $H^\pm \to  W^\pm V$
are depicted in Fig.~1. Those diagrams contains vertex diagrams
(Fig.~1.1 $\to$ 1.11), $W^\pm$-$H^\pm$ mixing (Fig.~1.12 $\to$ 1.14),
$H^\pm$-$G^\pm$ mixing (Fig.~1.15 $\to$ 1.17) and
$H^\pm$-$W^\pm$ mixing (Fig.~1.18 $\to$ 1.20).

Note that the mixing $H^\pm$--$W^\pm$ (Fig~.1.12, 1.13, 1.14) vanishes for 
an on-shell transverse W gauge boson.  
There is no contribution from the $W^\pm$--$G^\mp$ 
mixing because $\gamma G^\pm H^\mp$ and Z$G^\pm H^\mp$ vertices 
are absent at the tree level. 
All the Feynman diagrams have been generated and computed using 
FeynArts and FormCalc \cite{seep} packages.
We also used  the fortran FF--package \cite{ff} in the numerical 
analysis.

Although the amplitude for our process is absent at the tree level,
complications like tadpole contributions and 
vector boson--scalar mixing require a careful treatment of 
renormalization. We adopt, hereafter, the on-shell renormalization 
scheme of \cite{dabelstein}, for the Higgs sector, which is an 
extension of the on-shell scheme in \cite{Hollik}. 
In this scheme,  field renormalization is performed in the 
manifest-symmetric version of the Lagrangian. 
A field renormalization constant $Z_{\Phi_{1,2}}$ is assigned to each 
Higgs doublet 
$\Phi_{1,2}$. Following the same approach adopted in \cite{achm},
the Higgs fields and vacuum expectation values
 $v_i$ are renormalized as follows:    
\begin{eqnarray}
& &\Phi_i \rightarrow (Z_{\Phi_i})^{1/2} \Phi_i \nonumber\\
& &v_i \rightarrow (Z_{\Phi_i})^{1/2} (v_i-\delta v_i) \, . 
\label{reno}
\end{eqnarray}
With these substitutions in the scalar covariant derivative 
Lagrangian of the Higgs fields (in the convention of \cite{HHG}), 
followed by expanding the renormalization constants $Z_i=1+\delta Z_i$ 
to  the one-loop order, we obtain all the counter-terms 
relevant for our process:
\begin{eqnarray}
& &\delta [W_\nu^\pm H^\mp] \quad =  \, k^{\mu} \Delta 
\label{deltac} \\
& &\delta [A_\nu W_\mu^\pm H^\mp] = -i e g_{\mu\nu} \Delta
\label{deltaa} \\
& &\delta [Z_\nu W_\mu^\pm H^\mp]\, = -i e g_{\mu\nu}
\frac{s_W}{c_W}
\Delta  \label{deltab}
\end{eqnarray} 
where $k$ denotes the momentum of the incoming $W^\pm$ and
\begin{equation}
\label{deltadef}
   \Delta =\frac{\sin 2 \beta}{2} m_W
    [\frac{\delta v_2}{v_2} - \frac{\delta v_1}{v_1} +
     \delta Z_{\Phi_1} - \delta Z_{\Phi_2}  ] \, . 
\end{equation}
Denoting the one particle irreducible (1PI) two
point function for $W^\pm H^\pm$ (resp $G^\pm H^\pm$) mixing by 
$\pm ik_{\mu} \Sigma_{W^\pm H^\pm}(k^2)$ 
(resp $i \Sigma_{G^\pm H^\pm}(k^2)$) where $k$ is the momentum of
the incoming $W^\pm$ (resp $G^\pm$), and $H^\pm$ is outgoing. 
The renormalized mixing will be denoted by $\hat\Sigma$.

In the on-shell scheme, we will use the following renormalization conditions:
\begin{itemize}
\item  The renormalized tadpoles, i.e. the sum of
      tadpole diagrams $T_{h,H}$ and tadpole counter-terms
      $\delta_{h,H}$ vanish:   
\[  T_{h} +\delta t_h=0, \quad T_H +\delta t_H=0 \, . \]
These conditions guarantee that $v_{1,2}$ appearing in the renormalized 
Lagrangian ${\cal L_R}$ are located at the minimum of the one-loop potential.
\item The real part of the renormalized 
non-diagonal self-energy $\hat{\Sigma}_{H^\pm W^\pm}(k^2)$ 
vanishes for an on-shell charged Higgs boson: 
\begin{eqnarray}
& & \Re e \hat{\Sigma}_{H^\pm W^\pm} (m_{H^{\pm}}^2)  =0   
\label{WH0}
\end{eqnarray}
This renormalization condition determines the term $\Delta$ to be
\begin{eqnarray}
& & \Delta= \Re e {\Sigma}_{H^\pm W^\pm} (m_{H^{\pm}}^2)   
\label{Del}
\end{eqnarray}
and consequently 
$\delta [A_\nu W_\mu^\pm H^\mp]$ 
and $\delta [Z_\nu W_\mu^\pm H^\mp]$ are also fixed.
\end{itemize} 
The last renormalization condition is sufficient 
to discard the real part of the $H^\pm$--$G^\pm$ mixing contribution
as well. Indeed,  using the Slavnov--Taylor identity \cite{9607485}   
\begin{eqnarray}
k^2 {\Sigma}_{H^\pm W^\pm}(k^2) -m_W {\Sigma}_{H^\pm G^\pm }(k^2)=0\ \quad
\mbox{at}\ \ \ k^2 = m_{H^\pm}^2 \label{ward}
\end{eqnarray}
which is valid also for the renormalized quantities 
together with eq.~(\ref{WH0}),  it follows that
\begin{eqnarray}
\Re e\hat{\Sigma}_{H^\pm G^\pm}(m_{H^\pm}^2)=0 \, . 
\end{eqnarray}
In particular, the Feynman diagrams depicted in Fig.~1.9 will not 
contribute with the above renormalization conditions, being purely 
real valued. 

To make the amplitude of Fig.1 Ultra-Violet finite we need to add the following
counter-terms:  counter-terms for
 $\gamma W^{\pm}H^{\mp}$ and $ZW^{\pm}H^{\mp}$ vertices Fig.2.2a,
a counter-term for the $W^\pm$-$H^\mp$ mixing Fig.2.2b, 2.2d, and 
a counter-term for the $G^\pm$-$H^\mp$ mixing Fig. 2.2c.

\section{Numerics and discussions}
In our numerical evaluations, we use the following experimental 
input quantities~\cite{pdg4}: $\alpha^{-1}=129$, $m_Z$, $m_W$, $m_t$,
$m_b=$ 91.1875, 80.45, 174.3, 4.7  GeV. In the MSSM,
we specify the free parameters that will be used as follow: 
i) The MSSM Higgs sector is  parametrized by the CP-odd mass 
$m_{A^0}$ and $\tan\beta$, taking into account 
radiative corrections from \cite{sven}, and we assume  
$\tan\beta \ga 3$. ii) The chargino--neutralino sector can be 
parametrized by the  gaugino-mass  terms
$M_1$, $M_2$, and the Higgsino-mass term $\mu$. For simplification  
GUT relation $M_1\approx M_2/2$ is assumed. iii) Sfermions are characterized 
by a common soft-breaking sfermion mass 
$M_{SUSY} \equiv \widetilde{M}_L=\widetilde{M}_R$, $\mu$ the parameter and the
soft trilinear couplings for third generation scalar fermions 
$A_{t,b,\tau}$. For simplicity,  we will take $A_t=A_b=A_\tau$.

When varying the MSSM parameters, 
we take into account also the following constraints:
$i)$  The extra contributions to the $\delta\rho$ parameter from the 
Higgs scalars should not exceed the current limits from precision 
measurements \cite{pdg4}: $|\delta\rho|\la 0.003$.
$ii)$ $b \to s \gamma$ constraint. The present world average for 
inclusive $b\to s \gamma$ rate is \cite{pdg4}
${\cal B}(B \rightarrow X_s \gamma) = (3.3 \pm 0.4) \times 10^{-4}$. 
We keep the $B \rightarrow X_s
\gamma$ branching ratio in the 3$\sigma$ range of (2.1--4.5)$\times
10^{-4}$.  The SM part  of 
 $B \rightarrow X_s \gamma$ is calculated up to NLO using the 
expression given in \cite{NK}. While for the MSSM part, the Wilson coefficient
$C_7$ and $C_8$ are included at LO in the framework of MSSM with 
CKM as the only source of flavor violation and are taken from 
\cite{c7c8}. $iii)$ We will assume that all SUSY particles Sfermions
and charginos are heavier than about 100 GeV; for the light CP even 
Higgs we assume $m_{h^0}\ga 98$ GeV and $\tan\beta\ga 3$.


The total width of the charged Higgs is computed at tree level from
\cite{abdel2} without any QCD improvement for its fermionic decays 
$H^\pm\to \bar{f}f'$. The SUSY channels
like $H^+\to \widetilde{f}_i \widetilde{f}_j'$ and 
$H^+\to \widetilde{\chi}_i^0\widetilde{\chi}_j^+$ are
included when kinematicaly allowed.
\begin{figure}
\begin{tabular}{cc}
\resizebox{88mm}{!}{\includegraphics{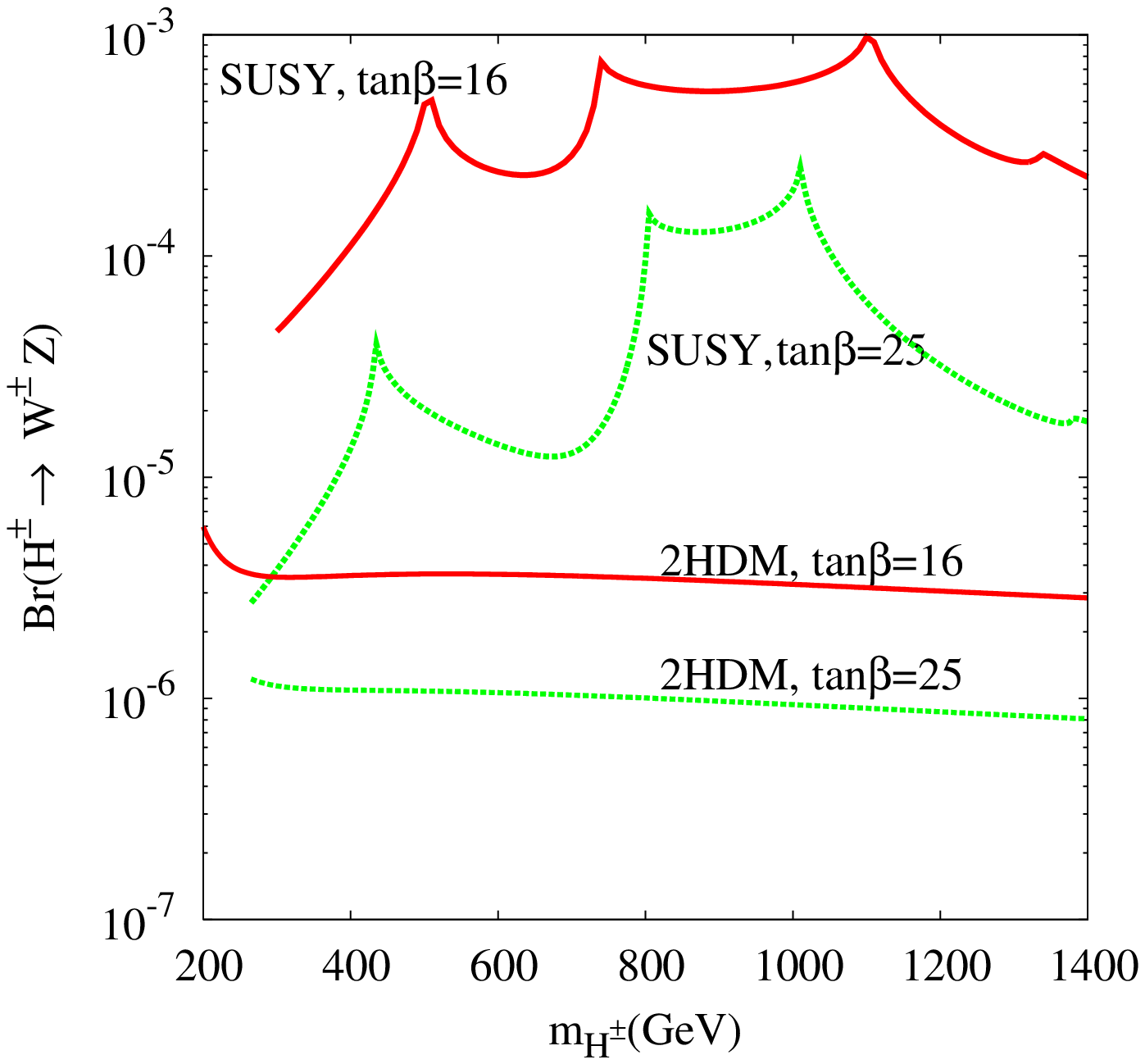}} &
\resizebox{88mm}{!}{\includegraphics{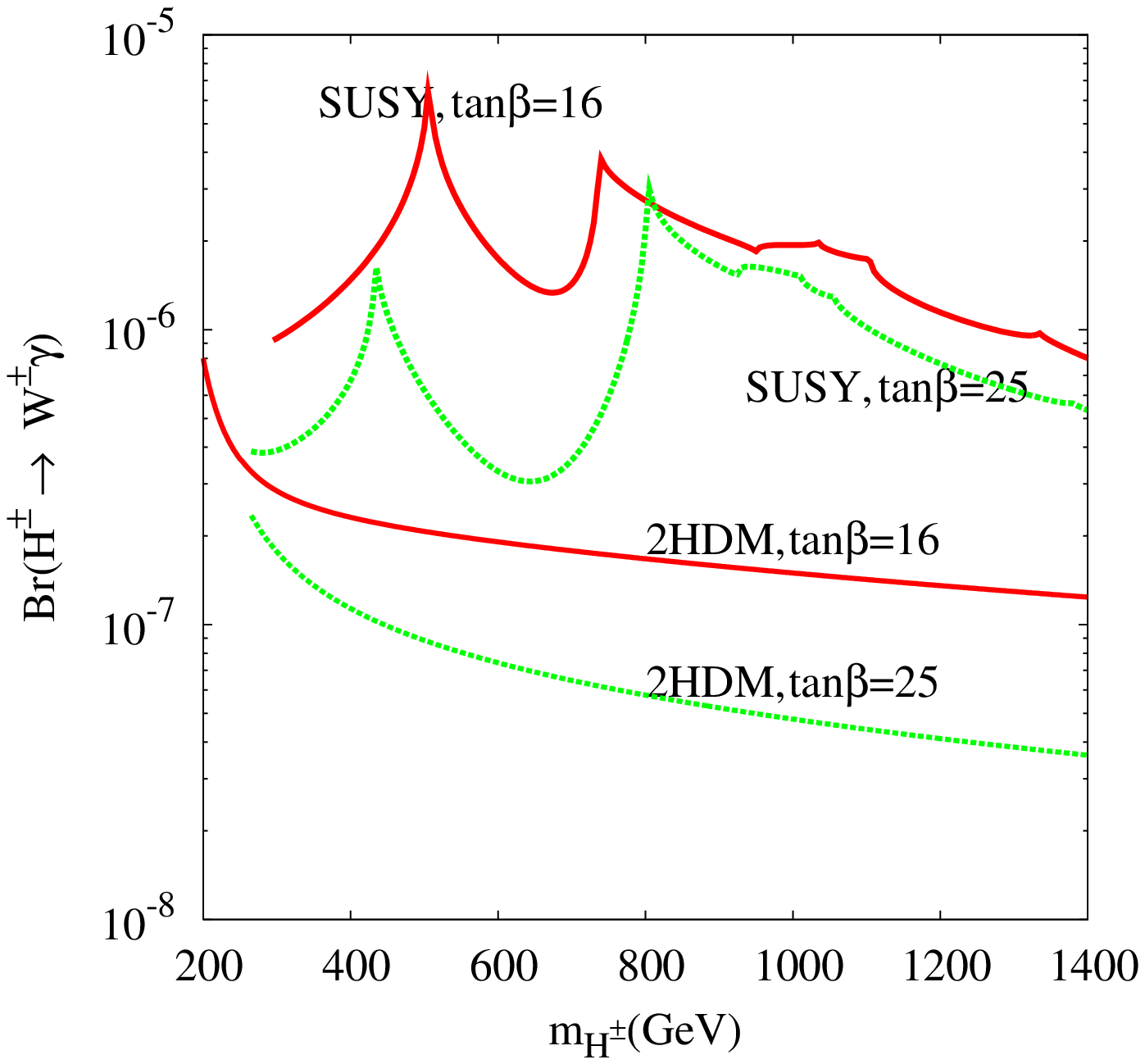}}
\end{tabular}
\caption
{Branching ratios of $H^{\pm}\to W^{\pm}Z$ (left) and 
$H^{\pm}\to W^{\pm}\gamma$ (right) as a function of 
$m_{H^{\pm}}$  in  the MSSM and 2HDM 
for $M_{SUSY}=500$ GeV, $M_2=175$ GeV, $\mu = -1.4$ GeV and
 $A_t =A_b = A_{\tau}= -\mu$ for various values of $\tan\beta$.}
\label{nfig1}
\end{figure}
\begin{figure}
\begin{tabular}{cc}
\resizebox{88mm}{!}{\includegraphics{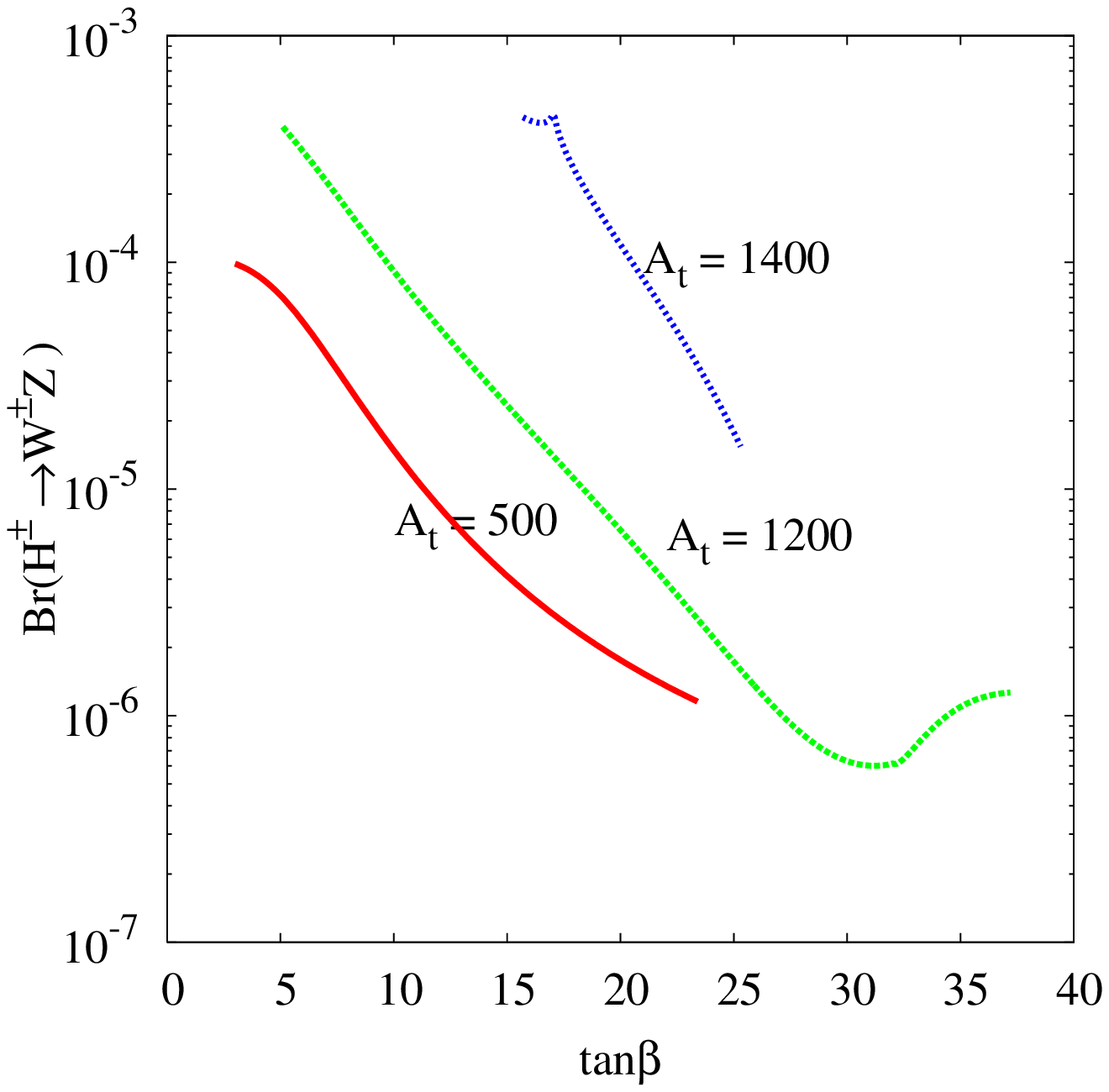}} &
\resizebox{88mm}{!}{\includegraphics{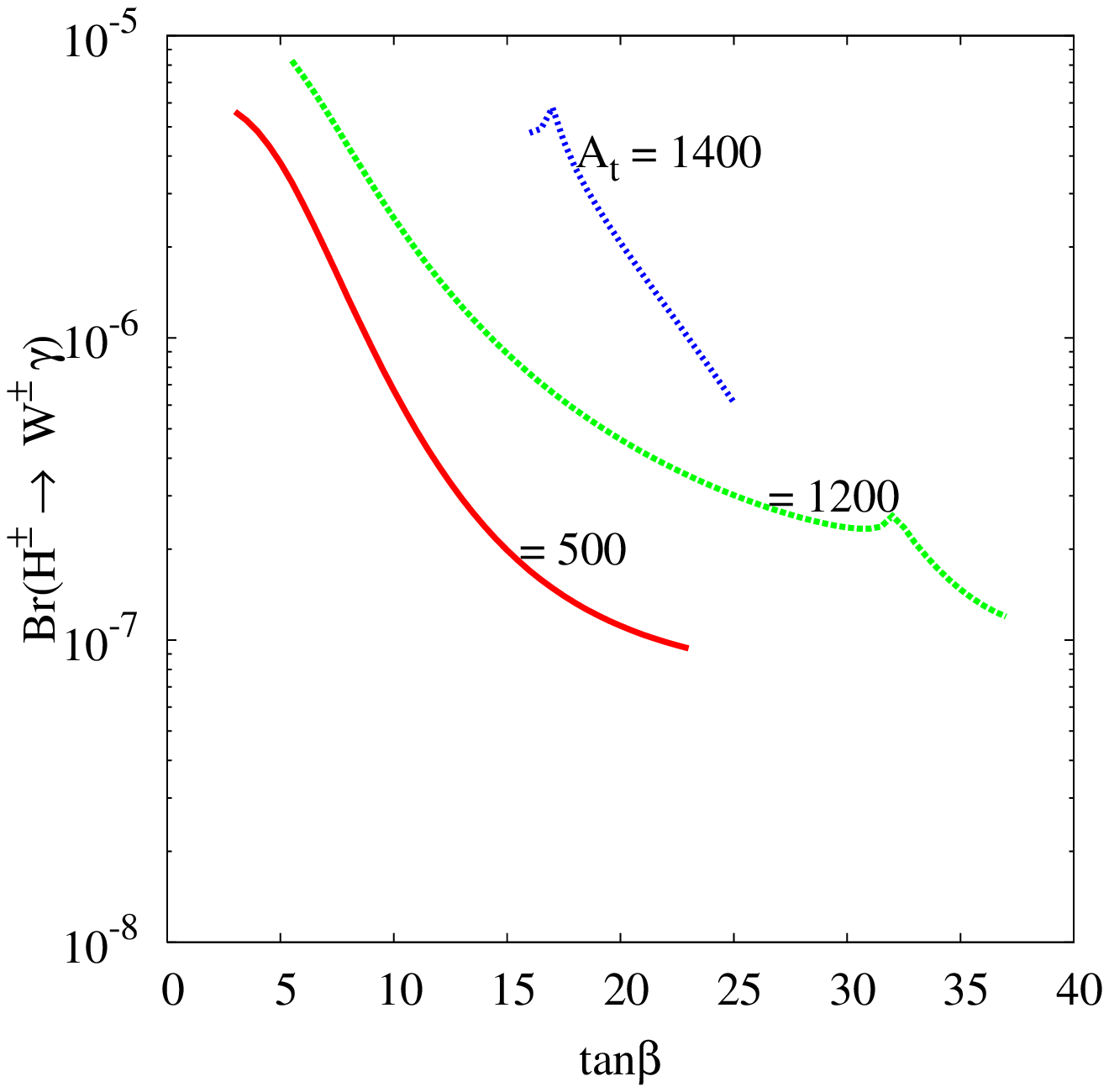}}
\end{tabular}
\caption
{Branching ratios for $H^{\pm}\to W^{\pm}Z$ (left), 
$H^{\pm}\to W^{\pm}\gamma$ (right) as a function of $\tan\beta$ 
in the MSSM with $M_{SUSY}=500$ GeV, 
$M_2=200$ GeV, $m_{H^{\pm}}=500$ GeV, 
$A_t = A_b=A_{\tau}=-\mu$ for various values of $A_t$.}
\label{nfig2}
\end{figure}
In Fig.~\ref{nfig1}, we show branching ratio of 
$H^{\pm}\to W^{\pm}Z$ (left) and  $H^{\pm}\to W^{\pm}\gamma$ (right)
as a function of charged Higgs mass for $\tan\beta=16$ and 25. 
In those plots, we have shown both the
pure 2HDM (only SM fermion, gauge bosons and Higgs bosons with 
MSSM sum rules for the Higgs sector) and the MSSM 
(2HDM and SUSY particles) contribution. As it can be seen from those plots,
both for $H^\pm \to W^\pm Z$ and $H^\pm\to W^\pm\gamma$ the 
2HDM contribution is rather small. Once we include the SUSY particles, 
we can see that the Branching fraction get enhanced and 
can reach $10^{-3}$ in case
of $H^\pm \to W^\pm Z$ and $10^{-5}$ in case of $H^\pm \to W^\pm \gamma$. 
The source of this enhancement is mainly due 
to the presence of scalar fermion contribution in the loop which are
amplified by threshold effects from the opening of the decay 
$H^\pm \to \widetilde{t}_i\widetilde{b}_j^*$. It turns out that the 
contribution of charginos neutralinos loops does not enhance the 
Branching fraction significantly as compared to scalar fermions loops.
The plots also show that, the branching fraction is more important for 
intermediate $\tan\beta=16$ and is slightly reduced for larger $\tan\beta=25$.

\begin{figure}
\begin{tabular}{cc}
\resizebox{88mm}{!}{\includegraphics{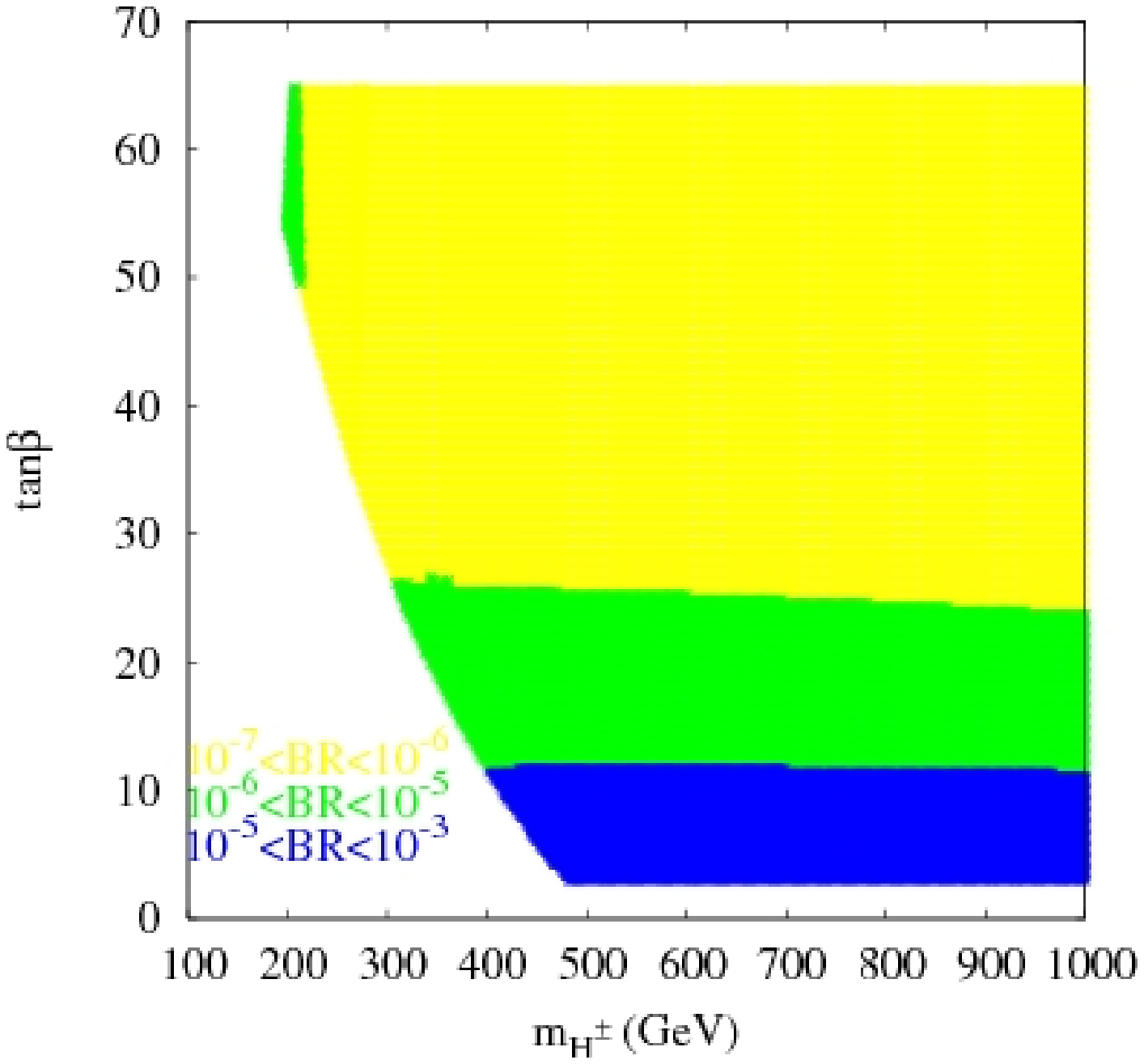}} &
\resizebox{88mm}{!}{\includegraphics{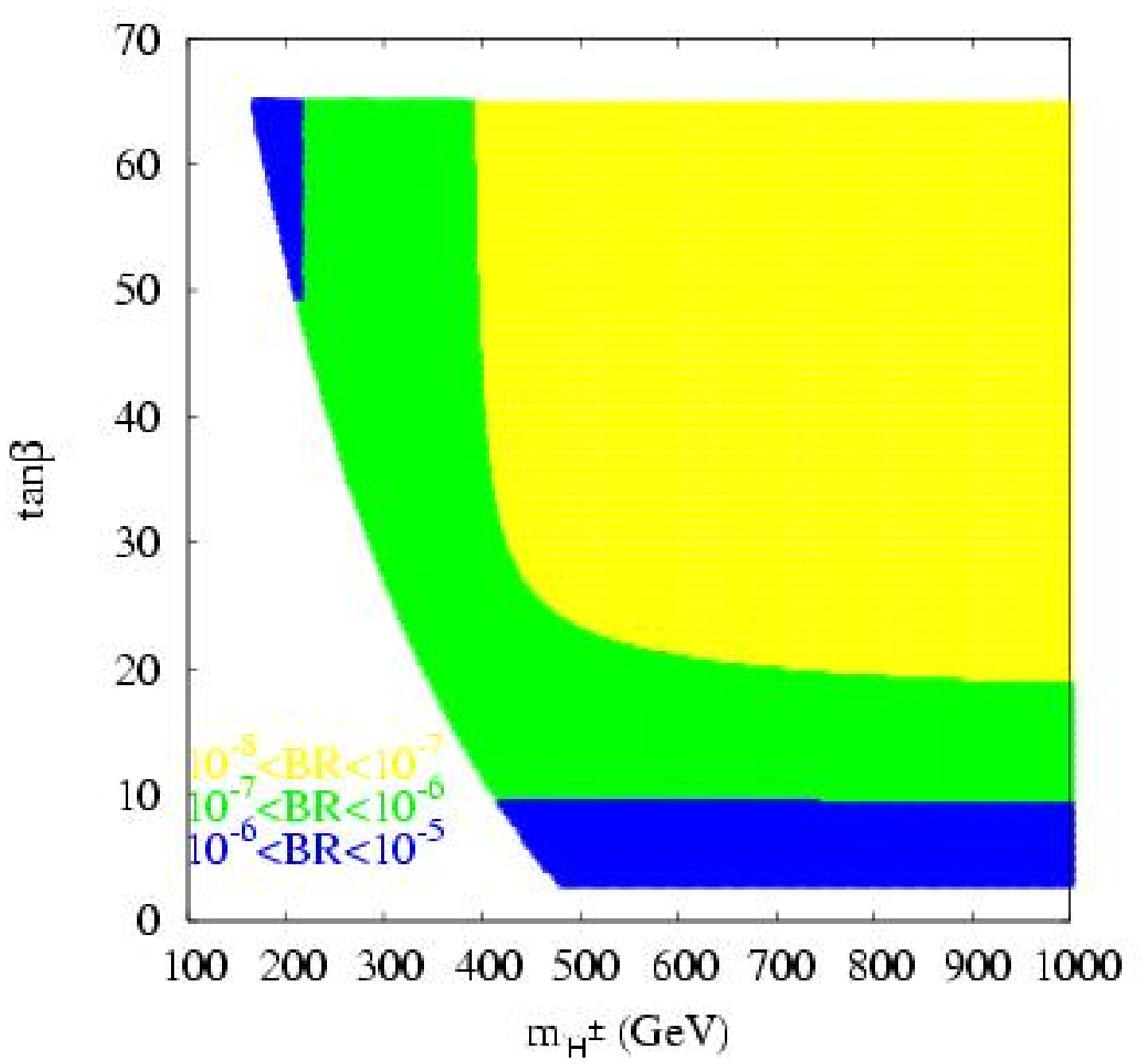}}
\end{tabular}
\caption
{Scatter plot for branching ratios of $H^{\pm}\to W^{\pm}Z$ 
 (left),$H^{\pm}\to W^{\pm}\gamma$ (right)
in the ($m_{H^{\pm}} $, $\tan\beta$) plane in the MSSM 
for $M_{SUSY} = 1$ TeV, $M_2 = 175$ GeV, 
$A_{t,b,\tau }=M_{SUSY}$ and $\mu = -1$ TeV.}
\label{nfig4}
\end{figure}

This $\tan\beta$ dependence is shown  in Fig.~\ref{nfig2}  both for 
$H^\pm\to W^\pm Z$ and $H^\pm \to W^\pm \gamma$ for three 
representative values of $A_t$.
It is obvious that the smallest is $\tan\beta$ the largest 
is the branching fraction. Increasing $\tan\beta$ from 5 to 
about 40 can reduce the branching fraction by about one or 
two order of magnitude.\\
We also show a scatter plot Fig.~\ref{nfig4} 
for $H^\pm\to W^\pm Z$ (left) and $H^\pm \to W^\pm \gamma$ (right)
 in ($m_{H^{\pm}} $, $\tan\beta$) plane for $A_t=-\mu=1$ TeV, 
$M_{SUSY}=A_t$ and $M_2=175$ GeV. As it can be seen from Fig~\ref{nfig4} 
there is only a small area for  $\tan\beta\la 10$ 
where the branching ratio of $H^{\pm}\to W^{\pm}Z$ can be in 
the range  $10^{-5}$--$10^{-3}$. 
\begin{figure}
\begin{tabular}{cc}
\resizebox{88mm}{!}{\includegraphics{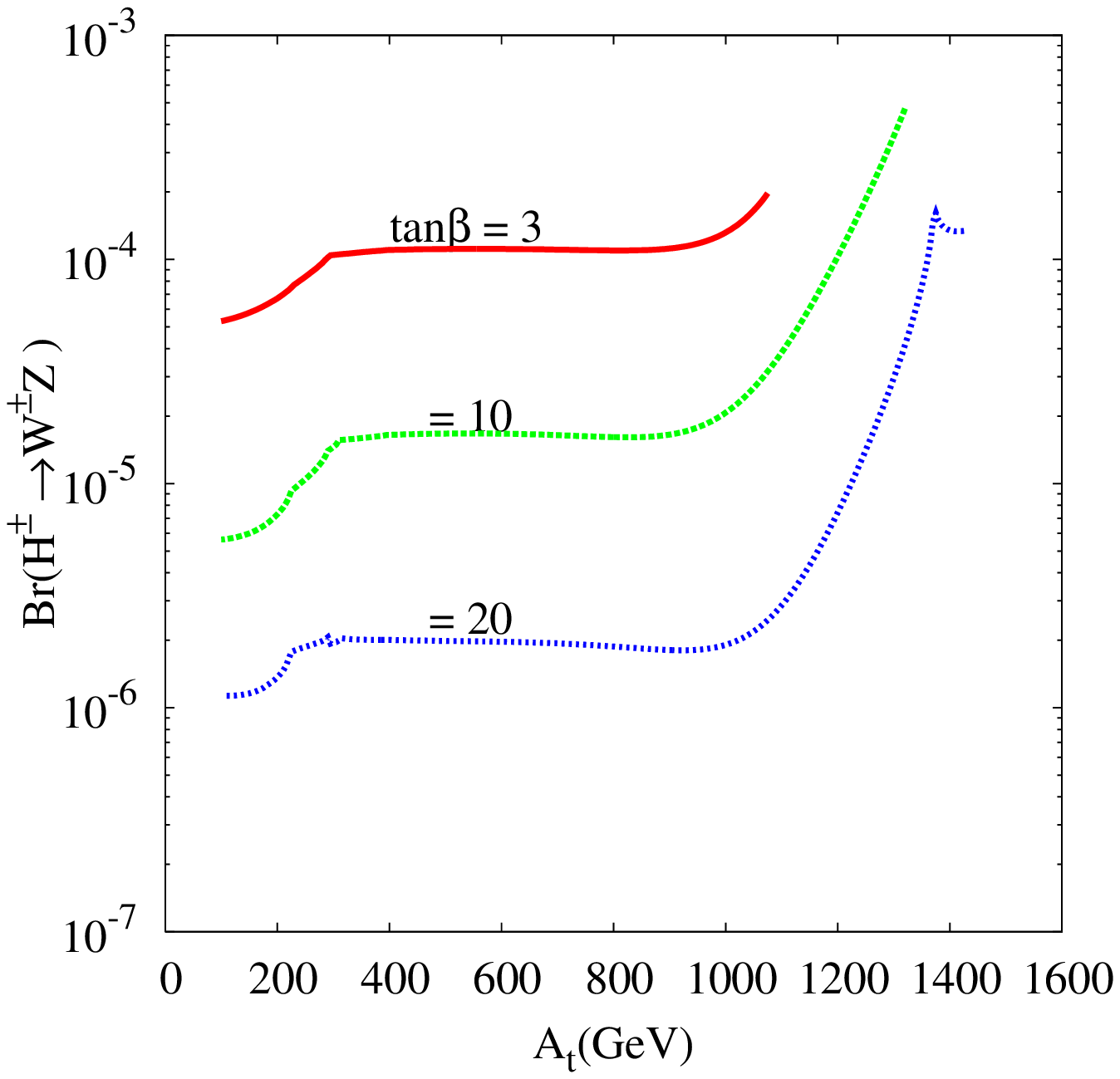}} &
\resizebox{88mm}{!}{\includegraphics{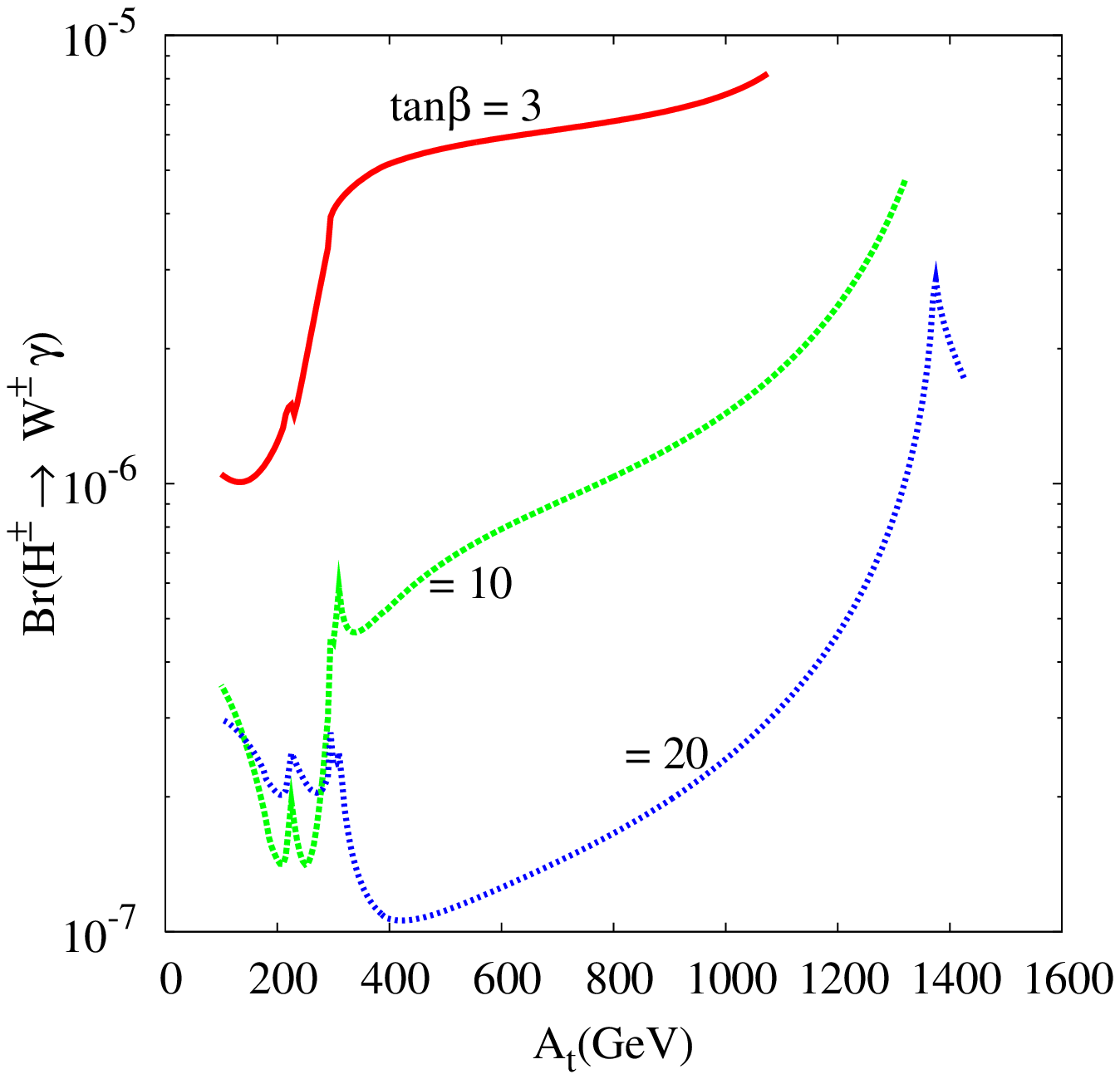}}
\end{tabular}
\caption
{Branching ratios for $H^{\pm}\to W^{\pm}Z$ (left) and
$H^{\pm}\to W^{\pm}\gamma$ (right) as a function of $A_t$  
in the MSSM with $M_{SUSY}=500$ GeV, $M_2=200$ GeV, 
$m_{H^{\pm}}=500$ GeV, $A_t = A_b= A_{\tau}= -\mu$ and $-2\, 
\rm{TeV} < \mu < -0.1$ TeV for various values of $\tan\beta$. }
\label{nfig3}
\end{figure}

We now illustrate in Fig.~\ref{nfig3} the branching fraction of 
$H^\pm\to W^\pm Z$ (left) and $H^\pm \to W^\pm \gamma$ (right) 
as a function of $A_t=A_b=A_{\tau}=-\mu $ for $M_{SUSY}=500$ GeV
and $M_2=200$ GeV. 
Since $b\to s \gamma$ favor $A_t$ and $\mu$ to have opposite sign,
we fix $\mu=-A_t$ and in this sense also $\mu$ is varied when 
$A_t$ is varied. Both for $H^\pm\to W^\pm Z$ and $H^\pm\to W^\pm \gamma$,
the chargino-neutralino contribution which is rather small
decrease with $\mu=-A_t$, the largest is $A_t$ the smallest is 
chargino-neutralino contribution.
In case of $H^\pm\to W^\pm Z$, for $A_t\la 1$ TeV it is 
the pure 2HDM contribution which dominate and that is why it is 
almost independent of $A_t$ while for large $A_t$ 
the branching ratio increase with $A_t$. 
It is clear that the largest is $A_t$ the 
largest is the branching ratio which can be of the order of $10^{-3}$ for
$H^\pm\to W^\pm Z$ with $\tan\beta=10$. 
As we know from $h^0\to \gamma \gamma$ and 
$h^0\to \gamma Z$ in MSSM \cite{whk},
the squarks contributions decouple except in the light stop mass 
and large $A_t$ limit \cite{whk}. 
In $H^\pm \to W^\pm V$ case, the same stuation happen.
As we can see from Fig.~\ref{nfig3} (left), for intermediate $A_t$,
$300<A_t<1000$ GeV, the squarks are rather heavy and hence their contributions
is small compared to 2HDM one. While for large $A_t$ 
the stop becomes very light $\la 200$ GeV and hence enhance 
$H^\pm \to W^\pm V$ width.
Of course this enhancement is also amplified by
$H^\pm\widetilde{t}_{L,R}\widetilde{b}_{R,L}^*$ and 
$H^\pm\widetilde{\tau}_{L,R}\widetilde{\nu_\tau}_L^*$ 
 couplings which are directly proportional to $A_{t,b,\tau}$. 
In case of $H^\pm\to W^\pm \gamma$ decay,
the pure 2HDM and sfermions contributon are of comparable size,
the branching ratio increases with $A_t$.

\begin{figure}
\begin{tabular}{cc}
\resizebox{88mm}{!}{\includegraphics{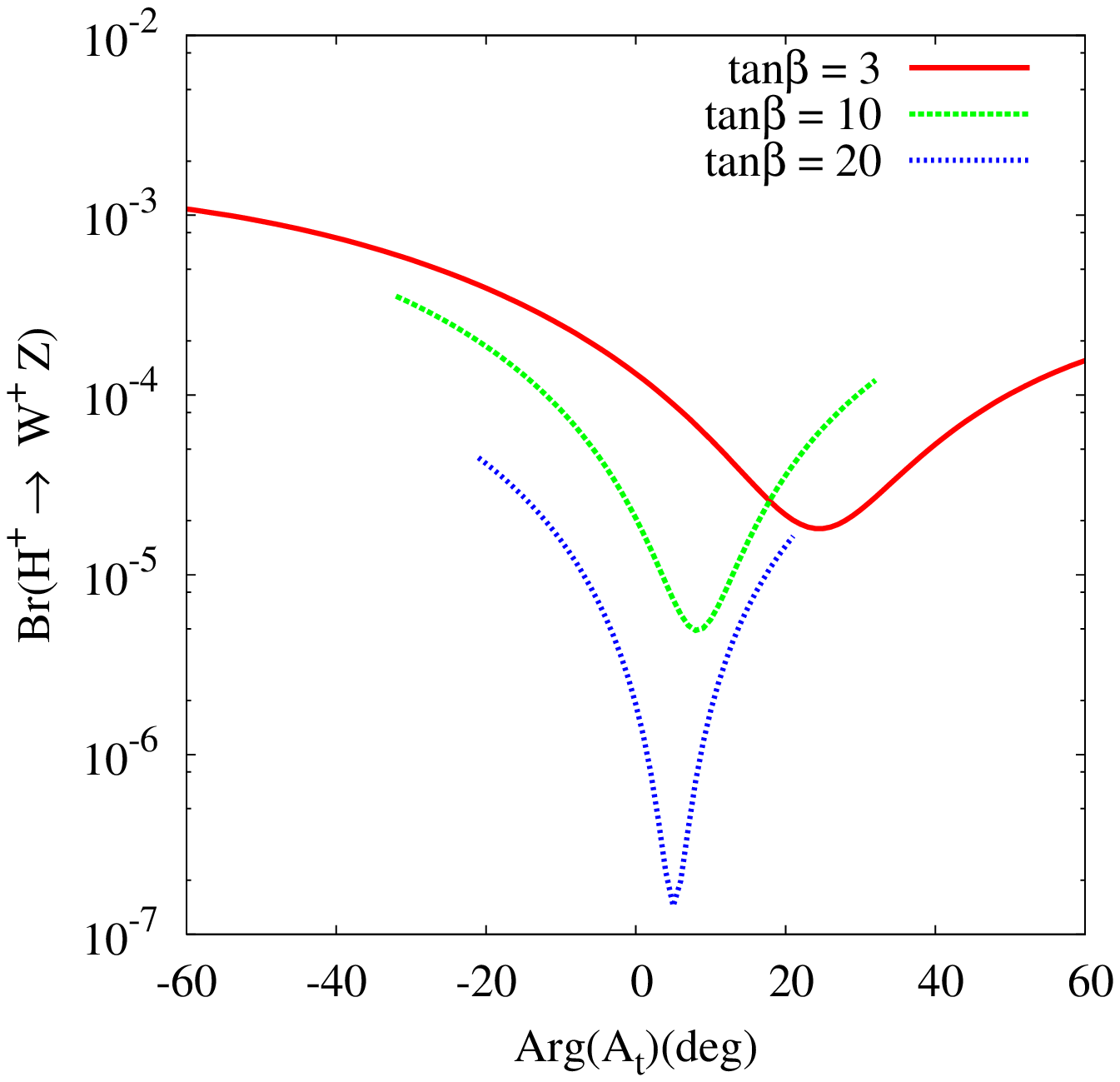}} &
\resizebox{88mm}{!}{\includegraphics{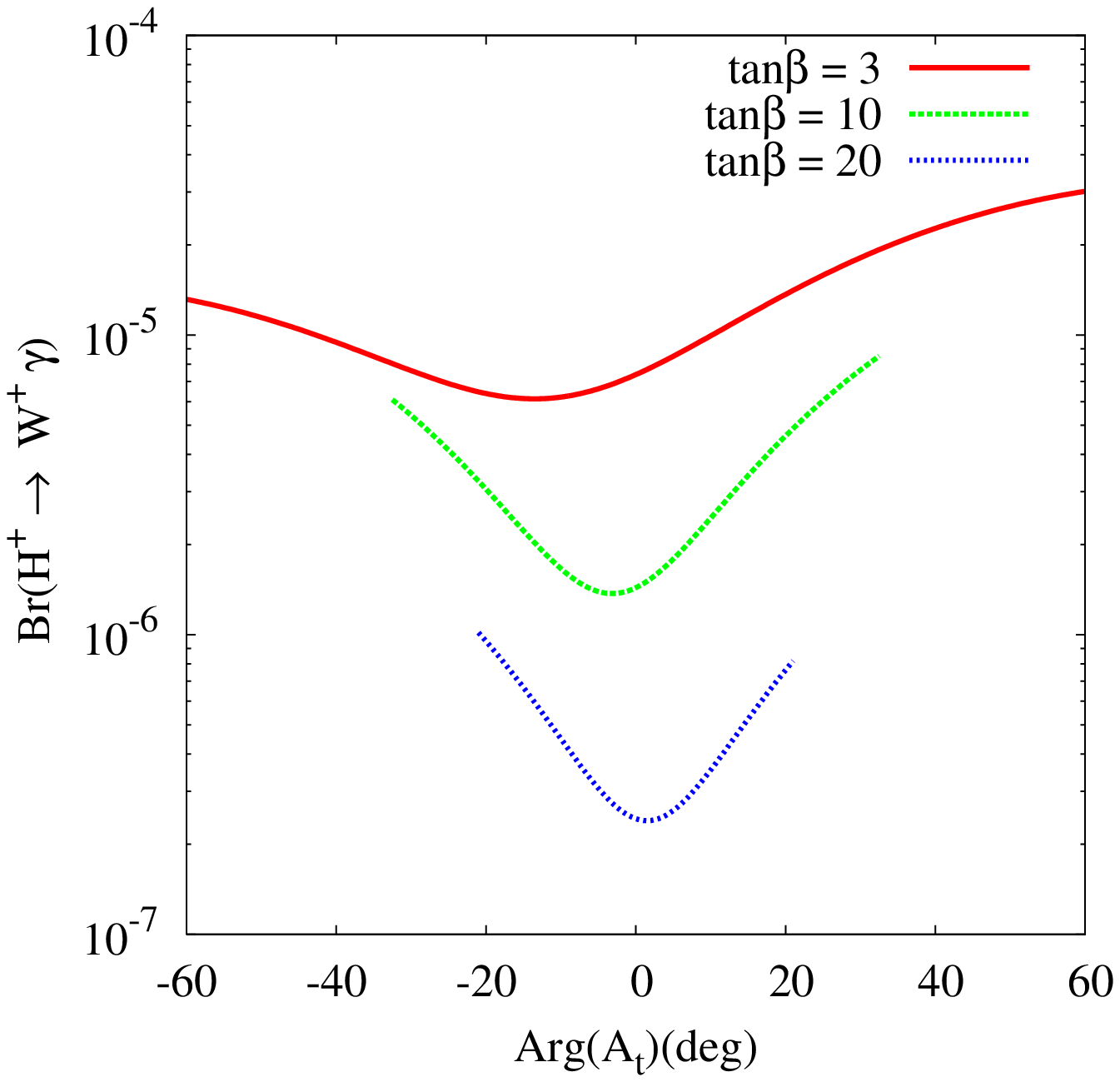}}
\end{tabular}
\caption
{Branching ratios for $H^{\pm}\to W^{\pm}Z$ (left) 
and $H^{\pm}\to W^{\pm}\gamma$ (right) in the
 MSSM as a function of $Arg(A_t)$ :
$M_{SUSY}=500$ GeV, $M_2=150\,GeV$, $m_{H^{\pm}}=500$ GeV, 
$A_t=A_b=A_{\tau}=-\mu =1$ TeV and for various values of $\tan\beta$.}
\label{figcp}
\end{figure}
We have also studied the effect of the MSSM CP violating phases.
It is well known that 
the presence of large SUSY CP violating phases can give contributions to 
electric dipole moments of the electron and neutron (EDM) 
which exceed the experimental upper bounds. In a variety of SUSY 
models such phases turn out to be severely constrained 
by such constraints i.e. ${\rm Arg}(\mu) < {\cal }(10^{-2})$ for a SUSY
mass scale of the order of few hundred GeV \cite{nath}.
For $H^\pm \to W^\pm Z$ and $H^\pm \to W^\pm \gamma$ decays
which are sensitive to MSSM CP violating phases through squarks and 
charginos-neutralinos contributions,
it turns out that the effect of MSSM CP violating phases is important
and can enhance  the rate by about one order of
magnitude. For illustration we show in Fig.~\ref{figcp} the effect
of $A_{t,b,\tau} $ CP violating phases for $M_{SUSY}=500$ GeV, 
$A_{t,b,\tau}=-\mu=$ 1 TeV and $M_2=150$ GeV. For simplicity,
we assume that $\mu$ is real. As it is clear,
the CP phase of $A_{t,b,\tau}$ can enhance the rates of both 
$H^\pm \to W^\pm Z, \gamma$ by more than an ordre of magnitude.
Those CP violating phases can lead to 
CP-violating rate asymmetry of $H^\pm$ decays,
those issues are going to be addressed in an incoming paper \cite{prep}.


We now turn to discuss the pure 2HDM contribution to 
$H^\pm\to W^\pm Z,W^\pm \gamma$. For the 2HDM parameterization we 
follow closely the notation of \cite{tch}.
In our discussion we will take as free parameters:
\begin{eqnarray}
\lambda_5 \quad , \quad m_{h^0} \quad , \quad m_{H^0} \quad , \quad
m_{A^0} \quad , \quad m_{H\pm} \quad , \quad \tan\beta \quad {\rm and} 
\quad \sin\alpha
\label{papa}
\end{eqnarray}
$\alpha$ is the CP-even mixing angle 
and $\lambda_5$ is the term that breaks softly 
the discrete symmetry
$\Phi_i\to -\Phi_i$ in the 2HDM Lagrangian.

To constrain the scalar sector parameters we will use both 
vacuum stability conditions as well as tree level unitarity and 
$\delta\rho$ constraints. In our study, we use the vacuum stability 
conditions from \cite{vac1} and tree level unitarity constraints 
are taken from \cite{kan}. \\
As $B\to X_s \gamma$ constraint is concerned, it has been shown 
in~\cite{bsg} that for models of the type 2HDM-II, data on $B\to X_s \gamma$ 
imposes a lower limit of $m_{H^\pm} \ga 350$\,GeV. In type I 2HDM, 
there is no such  constraint on the charged Higgs mass \cite{bsg}. 
In our numerical analysis we will ignore these constraints and allow 
$m_{H\pm}\la 200$ GeV in order to localize regions in parameter
space where the branching ratios are sizeable.

In the 2HDM, the source of enhancement in $H^\pm\to W^\pm Z$ and 
$H^\pm\to W^\pm \gamma$ decays can be either from bottom Yukawa coupling 
which is enhanced at large $\tan\beta$
or from the Higgs sector contribution through diagram like Fig.~1.2 with 
$(S_i,S_j,S_k)$=$(H^\pm ,h^0,H^\pm )$ or $(H^\pm ,H^0,H^\pm )$
and Fig.~1.8 with $(S_i,S_j)$=$(H^\pm ,h^0)$ or $(H^\pm ,H^0)$.
In contrast to the MSSM where the trilinear scalar couplings
$h^0H^+H^-$ and $H^0H^+H^-$ are function of the gauge couplings only,
in the 2HDM those couplings are function of Higgs masses, 
$\alpha$, $\tan\beta$ as well as $\lambda_5$ as it can be seen from 
their analytic expressions (in the notations of Ref.~\cite{tch})
\begin{eqnarray}
 {H^0H^+H^-}= & &
\frac{-ig}{m_W \sin 2\beta } (
m_{H^0}^2 (\cos^3\beta \sin\alpha +\sin^3\beta \cos\alpha)+
m_{H^{\pm}}^2\sin{2\beta} \cos({\beta-\alpha}) -
\sin({\beta+\alpha})\lambda_5 v^2 )\nonumber\label{scalar1}  \\
{h^0H^+H^-}  = && \frac{- ig}{m_W \sin 2\beta} ( 
 m_{h^0}^2(\cos{\alpha}\cos^3{\beta}- \sin{\alpha}\sin^3{\beta})
 +m_{H^{\pm}}^2\sin{2\beta}\sin({\beta-\alpha})  -
\cos({\beta+\alpha}){\lambda_5} v^2)\label{scalar3}  
\end{eqnarray}
Even after imposing unitarity and vacuum stability constraints,
those couplings can get large values compared to their MSSM values
and that is the main difference between 2HDM scalar sector and 
the MSSM one.

\begin{figure}
\begin{tabular}{cc}
\resizebox{88mm}{!}{\includegraphics{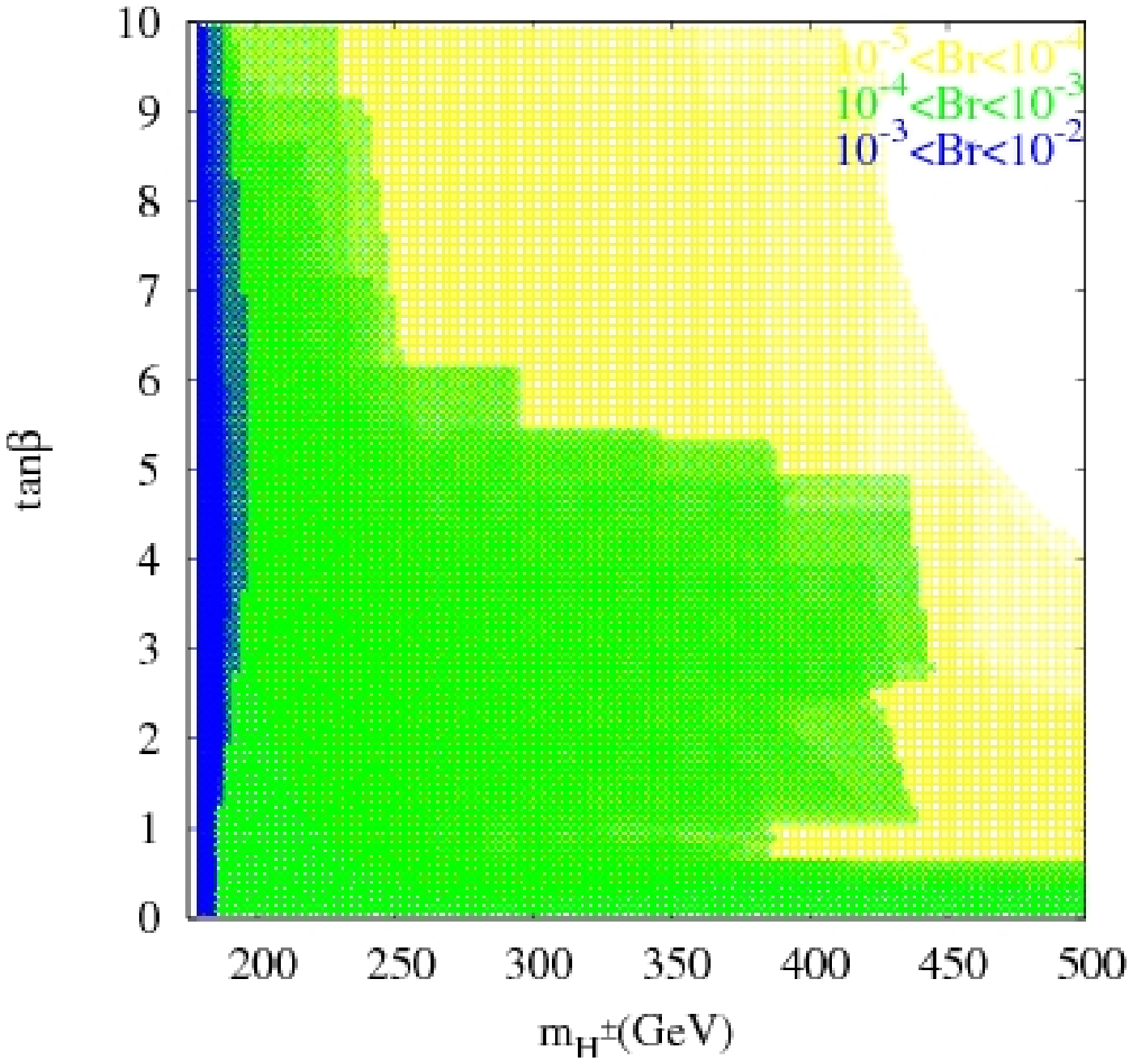}} &
\resizebox{88mm}{!}{\includegraphics{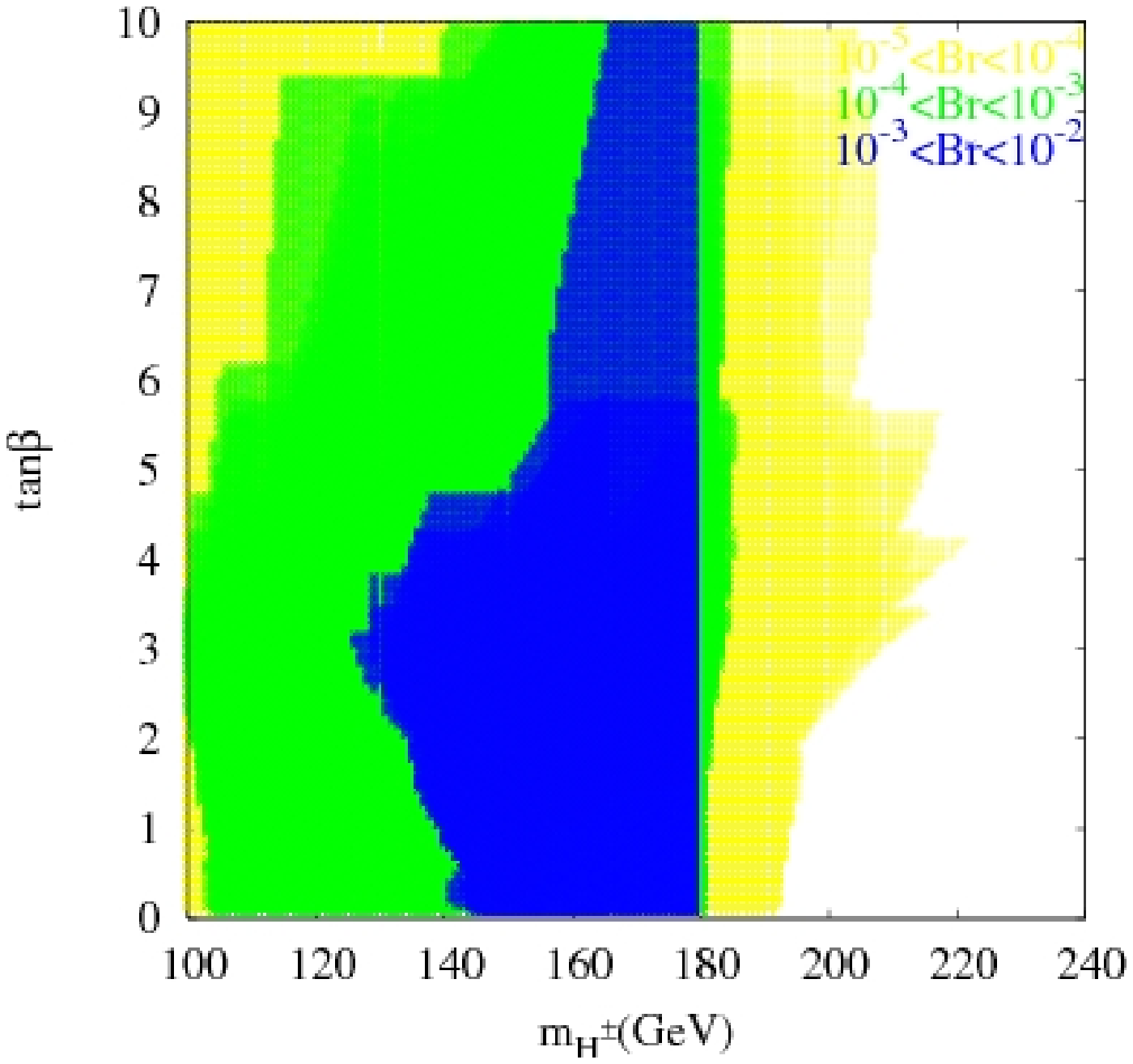}}
\end{tabular}
\caption
{The branching ratios of $H^{\pm}\to W^{\pm}Z$ (left) and
$H^{\pm}\to W^{\pm}\gamma$ (right) in type--II 2HDM model 
 in ($m_{H^{\pm}},\tan\beta$) plane for 
$100\,\rm{GeV}\le m_{h^0}\le\, 130\, \rm{GeV}$, 
$150\, \rm{GeV}\le\, m_{H^0}\le\, 350 \,\rm{GeV}$,
 $100\, \rm{GeV}\le m_{A^0,H^{\pm}}\le 500$ GeV, $-1\le\sin\alpha\le 1$, 
$0.1\le \tan\beta\le 10$, and $\lambda_5 = 0$. }
\label{2hdm1}
\end{figure}
\begin{figure}
\begin{tabular}{cc}
\resizebox{88mm}{!}{\includegraphics{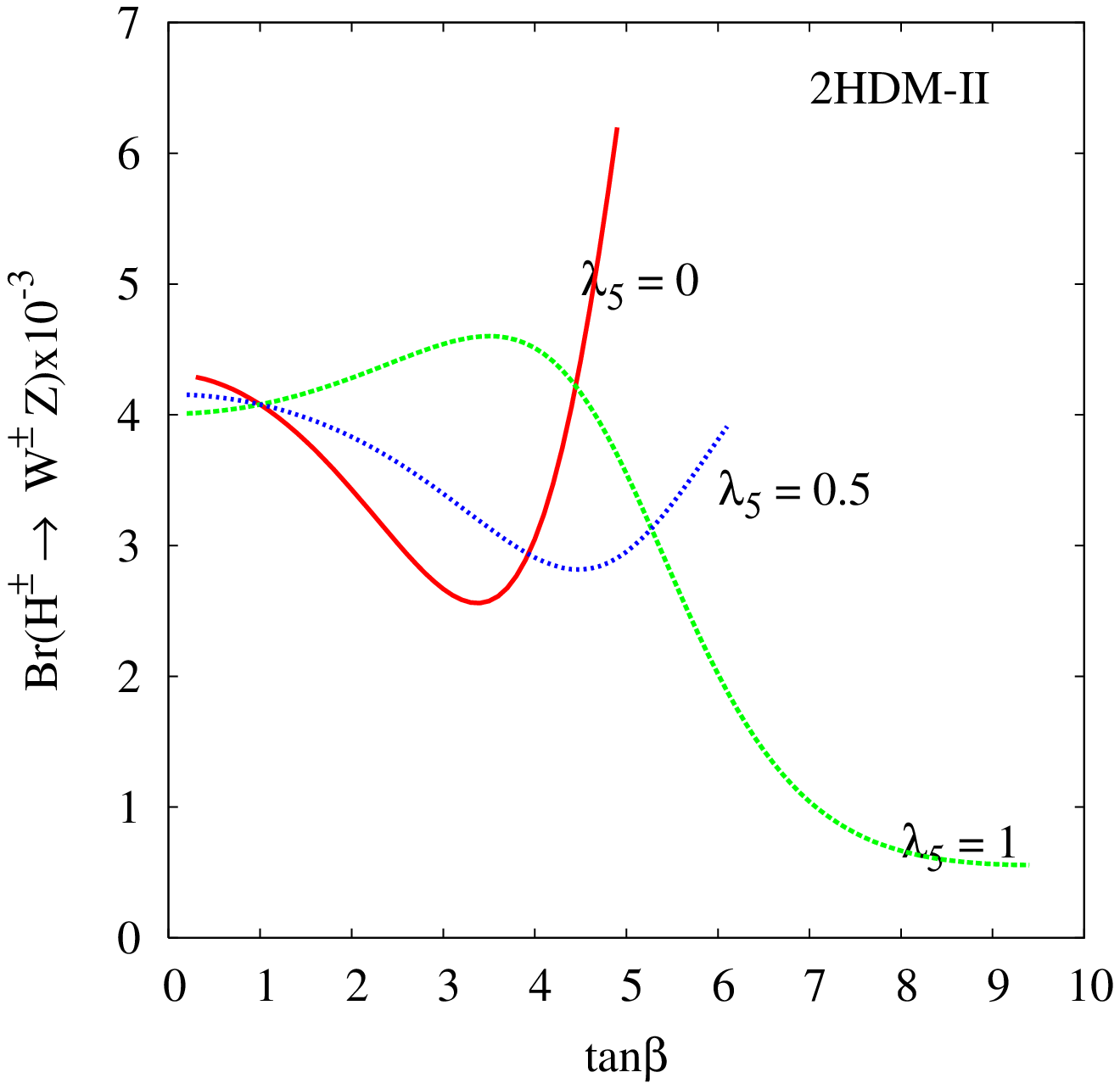}} &
\resizebox{88mm}{!}{\includegraphics{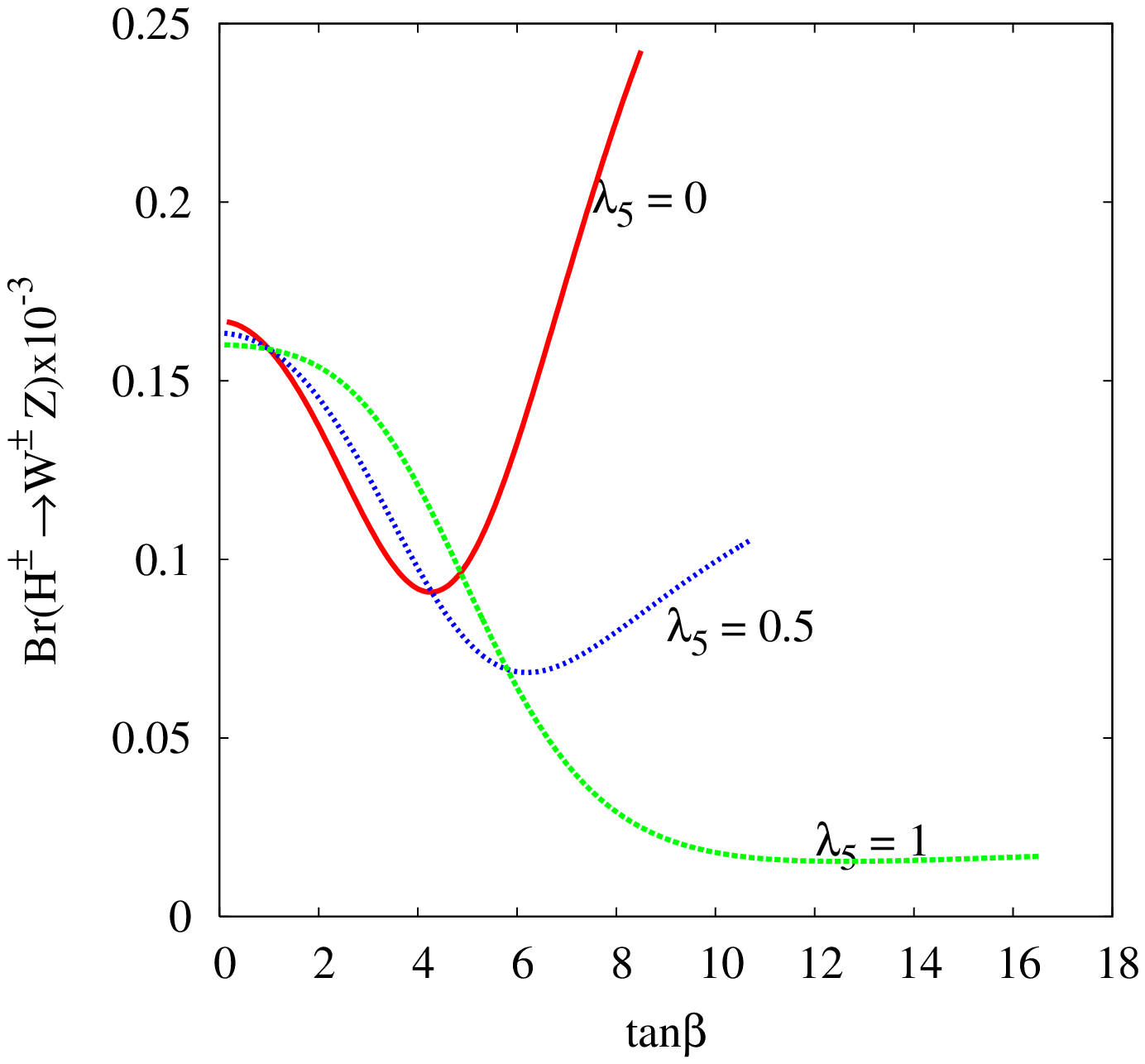}}
\end{tabular}
\caption
{Branching ratios $H^{\pm}\to W^{\pm}Z$ as the function of the $\tan\beta$ in
2HDM-II for $m_{h^0}=100$ GeV,
$m_{H^0}=200$ GeV, $m_{A^{0}}=300$ GeV
 $\alpha=\beta-\pi/2$ and various values of $\lambda_5$. Left plot is for 
$m_{H^{\pm}} = 180$ GeV and right one is for $m_{H^{\pm}} = 220$ GeV }  
\label{2hdm2}
\end{figure}

We have first compared our results to previous one given in 
Refs.~\cite{kanemur,toscano} and found perfect agreement. However
in Refs.~\cite{kanemur,toscano}, unitarity constraints were not imposed
while Ref.~\cite{toscano} did not consider the case of $H^\pm \to W^\pm Z$.
We have performed a systematic scan over the 2HDM parameters and found that 
both $H^\pm \to W^\pm Z$ and $H^\pm \to W^\pm \gamma$ can reach $10^{-2}$
branching ratio.
In Fig.\ref{2hdm1} we have illustrated our results as scatter plot
in the $(\tan\beta, m_{H^\pm})$ plane. As one can see from the plot,
just before the opening of $H^- \to \bar{t}b$ decay,
for $m_{H^\pm}\la m_t+m_b$, the branching ratio of 
$H^\pm \to W^\pm Z$ and $H^\pm \to W^\pm \gamma$ can be of
the order $10^{-2}$--$10^{-3}$. Once the decay $H^- \to \bar{t}b$ 
is open for $m_{H^\pm}\ga m_t+m_b$, the charged Higgs width becomes large
and the branching ratio of 
$H^\pm \to W^\pm Z$ and $H^\pm \to W^\pm \gamma$ are suppressed. 
However as one can see, there is still a large area in the 
$(\tan\beta, m_{H^\pm})$ plane where the branching ratio of 
$H^\pm \to W^\pm Z$ and $H^\pm \to W^\pm \gamma$ can be larger than $10^{-4}$.

We illustrate in Fig.~\ref{2hdm2} the branching ratio of $H^\pm \to W^\pm Z$
as a function of $\tan\beta$ for various values of $\lambda_5$ and for 
$m_{H^{\pm}} = 180$ GeV (left plot) and $m_{H^{\pm}} = 220$ GeV (right plot).
It is clear from both plots that the branching ratio is 
slightly enhanced for vanishing $\lambda_5$ case, this effect 
has been noticed also by \cite{kanemur}. In the left plot
we are very close to $H^- \to \bar{t}b$ threshold, the branching ratio
is larger than $10^{-3}$. Away from the  threshold $H^- \to \bar{t}b$
and for $m_{H^{\pm}} = 220$ GeV (right plot) the branching ratio 
of $H^\pm \to W^\pm Z$ is slightly reduced to the level of $10^{-4}$.

\section{Conclusion}

In the framework of MSSM and 2HDM we have studied
charged Higgs decays into a pair of gauge bosons namely:
$H^\pm\to W^\pm Z$ and $H^\pm\to W^\pm \gamma $.
In the MSSM we have also studied the effects of MSSM CP violating phases.
In contrast to previous studies, we have performed the calculation in
the 'tHooft-Feynman gauge and used a renormalization  prescription to deal
with tadpoles, $W^\pm$--$H^\pm$ and $G^\pm$--$H^\pm$ mixing.
The study has been carried out taking into account the experimental 
constraint on the $\rho$ parameter, 
$b\to s \gamma$ constraint, unitarity constraints, 
and vacuum stability conditions
 on all scalar quartic couplings $\lambda_i$ in 2HDM case.
Numerical results for the branching ratios have been presented.
In the MSSM, we have shown that the  branching ratio
of $H^\pm \to W^\pm Z$ can reach $10^{-3}$ in some cases 
while $H^\pm \to W^\pm \gamma$ never exceed $10^{-5}$.
The effect of MSSM CP violating phases is also found to be important.\\
In the 2HDM we emphasize the effect coming from the 
pure trilinear scalar couplings such as $h^0H^+H^-$ and $H^0H^+H^-$.
We have shown that in 2HDM both $H^\pm \to W^\pm Z$ and 
$H^\pm \to W^\pm \gamma$ can have a branching ratio in the 
range $10^{-2}$--$10^{-3}$. \\
Those Branching ratio in the range $10^{-2}$--$10^{-3}$ might provide 
an opportunity to search for a charged Higgs boson at the LHC 
through $H^\pm \to W^{\pm} Z$.

\acknowledgements 
A.A is grateful to Max-Planck Institute Munich for their kind 
hospitality during his visit where part of this work has been done.
This work is supported by PROTARS-III D16/04.

\appendix
\section{Lagrangian and couplings}
In this appendix, we list our notation for the gauge and Yukawa
couplings  used in this section.

The photon coupling  to fermions/sfermions are:
\begin{equation}
g_{\gamma ff}^{L,R}=-e_f\sw, \ \ \ g_{\gamma\tlf_i\tlf_j}=-e_f\sw\delta_{ij}
\end{equation}

The $Z$ boson coupling  to fermions/sfermions are:
\begin{eqnarray}
g_{Zff}^{L}&=&(-T_3+e_f\sw^2)/c_W,\ \ \ g_{Zff}^{R}=e_f\sw^2/c_W\\
g_{Z\tlf_i\tlf_j}&=&c_{ij}/\cw
\end{eqnarray}
where the couplings $c_{ij}$ are given by
 \begin{equation}
c_{ij}=\left(
\begin{array}{cc}
T_3\cos^2_{\theta_f}-e_f\sw^2&-\frac{1}{2}T_3\sin2\theta_f\\
-\frac{1}{2}T_3\sin2\theta_f&T_3\sin^2_{\theta_f}-e_f\sw^2
\end{array}
\right)
\end{equation}
where $T_3=1/2$ for up quarks and $T_3=-1/2$ for down quarks and 
$\theta_f$ is the mixing angle in the sfermion sector defined by:
\begin{equation}
\left(
\begin{array}{c}
\tilde{f}_L\\\tilde{f}_R
\end{array}
\right) =
\left(
\begin{array}{cc}
\cat &-\sat \\
\sat& \cat 
\end{array}\right)
\left(\begin{array}{c}\tilde{f}_1\\\tilde{f}_2\end{array}\right)
=R^f_{ij}\left(\begin{array}{c}\tilde{f}_1\\\tilde{f}_2\end{array}\right)
\end{equation}
where $\tilde f_{L,R}$ are the weak eigenstates and $\tilde f_{1,2}$
are the mass eigenstates.

The $W$ gauge boson coupling to a pair of fermions/sfermions is: 
\begin{eqnarray}
g_{Wff^{\prime}}^{L}&=&-\frac{1}{\sqrt{2}},\ \ \ g_{Wff^{\prime}}^{R}=0\\
g_{W\tlf_i\tlf_j}&=&-\frac{1}{\sqrt{2}}R^{f}_{i1}R^{f^{\prime}}_{j1}
\end{eqnarray}

The chargino mass matrix is:
\begin{equation} 
X = \left( \begin{array}{cc}
	M_2 & \sqrt{2} m_W \sin\beta \\
	\sqrt{2} m_W \cos\beta & \mu 
	\end{array}
	\right),
\end{equation}
which is diagonalized by the unitary matrices $U$ and $V$ via
$V X^{\dagger} U^{\dagger} = M_D$.

The neutralino mass matrix is:
\begin{equation}
Y = \left( \begin{array}{cccc}
	M_1 & 0 & -m_Z \sw \cos\beta & m_Z \sw \sin\beta \\
	0 & M_2 & m_Z \cw \cos\beta & -m_Z \cw \sin\beta \\
	-m_Z \sw \cos\beta & m_Z \cw \cos\beta & 0 & -\mu \\
	m_Z \sw \sin\beta & -m_Z \cw \sin\beta & -\mu & 0
	\end{array} \right),
\end{equation}

which is diagonalized by the matrix $N$ via
$N Y^{\dagger} N^{\dagger} = N_D$.

The matrices that enter the $W^\pm\tilde\chi_i^0\tilde\chi_j^\mp$ 
couplings are defined as:
\begin{equation}
\cO_{ij}^L=-\frac{1}{\sqrt{2}}N_{i4}V_{j2}^*+N_{i2}V_{j1}^*,\ \ \ 
\cO_{ij}^R=\frac{1}{\sqrt{2}}N_{i3}^*U_{j2}+N_{i2}^*U_{j1}.
\end{equation}

The matrices that enter the $Z\tilde\chi_i^\pm\tilde\chi_j^\mp$ couplings 
are defined as:
\begin{eqnarray}
\cO_{ij}^{\prime{L}}&=&-V_{i1}V_{j1}^*-\frac{1}{2}V_{i2}V_{j2}^*
+\delta_{ij}\sw^2, \nonumber \\
\cO_{ij}^{\prime{R}}&=&-U_{i1}^*U_{j1}-\frac{1}{2}U_{i2}^*U_{j2}
+\delta_{ij}\sw^2.
\end{eqnarray}

The matrices that enter the $Z\tilde\chi_i^0\tilde\chi_j^0$ couplings 
are defined as:
\begin{equation}
\cO_{ij}^{\prime\prime{L}} = -\frac{1}{2}N_{i3}N_{j3}^*
+\frac{1}{2}N_{i4}N_{j4}^*, \ \ \ 
\cO_{ij}^{\prime\prime{R}} = -O_{ij}^{\prime\prime{L}*}.
\end{equation}

The matrices that enter the $H^{\mp}\tilde\chi_i^0\tilde\chi_j^\pm$ 
and $H^{\mp}u_jd_j$ couplings are defined as:
\begin{eqnarray}
\cO_{ij}^{\prime{L}}&=&N_{i4}^*V_{j1}^*
+\frac{1}{\sqrt{2}}(N_{i2}^*+N_{i1}^*\tan\theta_W)V_{j2}^*, \nonumber \\
\cO_{ij}^{\prime{R}}&=&N_{i3}U_{j1}-\frac{1}{\sqrt{2}}
(N_{i2}+N_{i1}\tan\theta_W)U_{j2}.\non\\
g_{H^\pm u_i d_i}^{L,R} &=& \frac{(m_{d_i}
\tan\beta,m_{u_i}/\tan\beta)}{\sqrt{2}m_W},\ \ \ 
g_{G^\pm u_i d_i}^{L,R} = \frac{m_{u_i},m_{d_i}}{\sqrt{2}m_W}\\\non
\end{eqnarray}
The couplings $W^\pm G^\mp V$, $W^\pm 
W^\mp V$ and $H^\pm H^\mp V $ are given: 
\begin{center}
\begin{tabular}{|c|c|c|c|} \hline
$V$&$g_{WGV}$& $g_{WWV}$&$g_{HHV}$\\ \hline
$Z$&$-m_Z\sw^2$&$\cw$&$(\cw^2-\sw^2)/2\cw$\\
$\gamma$&$\mw\sw$&$-\sw$&$\sw$\\
\hline
\end{tabular}
\end{center}
The coupling $H^\mp \tilde f \tilde f^{\prime}$ are given by
\begin{eqnarray}   
\ghenu &=& -\mw\sin{2}\beta/\sqrt{2} \nonumber\\
\ghdu&=&-\mw\sin{2}\beta/\sqrt{2}\nonumber\\
g_{H^-\sbl\stl}&=&-\mw\sin{2}\beta/\sqrt{2} 
         + (\mb^2\tan\beta + \mt^2\cot\beta)/\sqrt{2}\mw \nonumber\\
g_{H^-\sbl\str}&=&\mt(\mu + A_t\cot\beta)/\sqrt{2}\mw  \nonumber\\
g_{H^-\sbr\stl}&=&\mb(\mu + A_b\tan\beta)/\sqrt{2}\mw \nonumber\\
g_{H^-\sbr\str}&=&\mt\mb(\tan\beta+\cot\beta)/\sqrt{2}\mw 
\end{eqnarray}
\begin{eqnarray}
\ghbonetone&=&c_{\theta_t}c_{\theta_b}{g}_{H^-\sbl\stl}+s_{\theta_t}c_{\theta_b}{g}_{H^-\sbl\str}+c_{\theta_t}s_{\theta_b}{g}_{H^-\sbr\stl}+s_{\theta_t}s_{\theta_b}{g}_{H^-\sbr\str} \nonumber\\
\ghbonettwo&=&-s_{\theta_t}c_{\theta_b}{g}_{H^-\sbl\stl}+c_{\theta_t}c_{\theta_b}{g}_{H^-\sbl\str}-s_{\theta_t}s_{\theta_b}{g}_{H^-\sbr\stl}+c_{\theta_t}s_{\theta_b}{g}_{H^-\sbr\str}\nonumber \\
\ghbtwotone&=&-c_{\theta_t}s_{\theta_b}{g}_{H^-\sbl\stl}-s_{\theta_t}s_{\theta_b}{g}_{H^-\sbl\str}+c_{\theta_t}c_{\theta_b}{g}_{H^-\sbr\stl}+s_{\theta_t}c_{\theta_b}{g}_{H^-\sbr\str}\nonumber \\
\ghbtwottwo&=&s_{\theta_t}s_{\theta_b}{g}_{H^-\sbl\stl}-c_{\theta_t}s_{\theta_b}{g}_{H^-\sbl\str}-s_{\theta_t}c_{\theta_b}{g}_{H^-\sbr\stl}+c_{\theta_t}c_{\theta_b}{g}_{H^-\sbr\str} 
\end{eqnarray}
\section{MSSM contribution for $H^{\pm}\to W^{\pm}V$}
We now list our results for the MSSM diagrams.  
The matrix elements for $H^{\pm} \to W^{\pm} V$ ($V=Z,\gamma$) in the MSSM 
are partially presented here. We give only the fermionic contribution 
(SM fermion and chargino-neutralino contribution) as well as the scalar 
fermion contributions. For the bosonic contribution (higgs bosons and 
gauge bosons) we refer to \cite{kanemur}. 
\\
It is to be understood
that diagrams involving charginos $\tilde\chi^\pm_i$ are summed over $i=1,2$
and diagrams involving neutralinos $\tilde\chi^0_i$ are summed over
$i=1,\ldots,4$.\\
The fermion triangle that enters 
Fig.~\ref{hwz}.1a and Fig.~\ref{hwz}.1b was computed in Ref.~\cite{micapey}, 
and agrees with our result.
Fig.~\ref{hwz}.1c
with $\tilde\chi_i^0$, $\tilde\chi_j^0$ and $\tilde\chi_k^\pm$ in the loop 
is analogous to the top/bottom quark triangle diagram and can be
checked by substituting top/bottom quark couplings into the gaugino
couplings.

Diagram Fig.~\ref{hwz}.1a:
In the convention of eq.\ref{lor}, we have
\begin{eqnarray}
{\cal F}_1&=&N_C\glwtb\Big(2(\mt\glvbb\grhtb+\mb\glvbb\glhtb-\mb\grvbb\glhtb)B_{0}-2\mt\mb\grvbb(\mt\glhtb+\mb\grhtb)C_0\\\non&+&\Big[-\mb\grvbb\glhtb(\mhp^2+\mw^2-m^2_V)-\mt\glvbb\grhtb(-\mhp^2-3\mw^2+m^2_V)+2\mb\glvbb\grhtb\mw^2\Big]C_1\\\non&+&\Big[\mt\glvbb\grhtb(3\mhp^2+\mw^2-m^2_V)+\mb\glvbb\glhtb(\mhp^2+\mw^2-m^2_V)-2\grvbb\glhtb\mhp^2\mb\Big]C_2\\\non&-&4\glvbb(\mt\grhtb+\mb\glhtb)C_{00}\Big)\\
{\cal F}_2&=& -2N_Cg^L_W\Big(\mt\glvbb\grhtb\,C_0-(\mt\glvbb\grhtb-\mb\grvbb\grhtb)\,C_1 + 3\glvbb(\mt\grhtb+\mb\glhtb)\,C_2\Big)\\
 {\cal F}_3&=&2N_Cg^L_W\Big(\mt\glvbb\grhtb\,C_0+(\mt\glvbb\grhtb+\mb\grvbb\glhtb)\,C_1+\glvbb(\mt\grhtb+\mb\glhtb)\,C_2\Big)
\end{eqnarray}

Where $N_C=3$ for quarks and 1 for leptons.
The arguments of the Passarino-Veltman functions $B_i$ and $C_{i}$ are
$B_i(m^2_V,m^2_{b},m^2_{b})$, 
$C_i(m^2_W,m^2_V,\mhp^2,m^2_{t},m^2_{b},m^2_{b})$.\\

Diagram Fig.~\ref{hwz}.1b:

Similar to Fig.~\ref{hwz}.1(a) with the exchange of
\begin{eqnarray}
\mt \leftrightarrow \mb,\ \ \ \grvtt\leftrightarrow\grvbb,\ \ \ 
\glvtt\leftrightarrow\glvbb,\ \ \ \glhtb\leftrightarrow\grhtb, \ \ \ 
g^L_{W}\leftrightarrow g^{L*}_W
\end{eqnarray}

Diagram Fig.~\ref{hwz}.1c

\begin{eqnarray}
{\cal F}_1 &=& \Big(2\Big[m_{\chi_i}(g_1g_3g_5+g_2g_4g_6)+m_{\chi_j}(g_1g_3g_6+
g_2g_4g_5)-m_{\chi_k}(g_1g_4g_5+g_2g_3g_6)\Big]\,B_0\\\non&+&\Big[m_{\chi_i}
(-2g_1g_4g_5m_{\chi_i}m_{\chi_k}-2g_1g_4g_6m_{\chi_j}m_{\chi_k}+
2g_1g_3g_6m_{\chi_i}m_{\chi_j}+2g_1g_3g_5m_{\chi_i}^2\\\non&&+
g_1g_3g_5(\mhp^2+m_W^2-m_V^2))\Big]\,C_0+\Big[-m_{\chi_k}(g_1g_4g_5+
g_2g_3g_6)(\mhp^2+m_W^2-m_V^2)\\\non&&+m_{\chi_i}(g_1g_3g_5-g_2g_4g_6)
(-\mhp^2-3m_W^2+m_V^2)-2(g_1g_3g_6-g_2g_4g_5)m_{\chi_j}m_W^2\Big]\,C_1\\\non&+
&\Big[m_{\chi_i}(g_1g_3g_5+g_2g_4g_6)(3\mhp^2+m_W^2-m_V^2)+
m_{\chi_j}(g_2g_4g_5+g_1g_3g_6)(\mhp^2+ m_W^2-m_V^2)\\\non&&
-2(g_1g_4g_5-g_2g_3g_5)m_{\chi_k}\mhp^2\Big]\,C_2\\\non&-&
4\Big[m_{\chi_i}(g_1g_3g_5+g_2g_4g_6)+m_{\chi_j}(g_1g_3g_6+g_2g_4g_5)\Big]\,
C_{00}\Big)\\
{\cal F}_2 &=& -2\Big((m_{\chi_i}(g_1g_3g_5 + g_2g_4g_6))\,C_0- 
\Big[m_{\chi_k}(g_2g_3g_6-g_1g_4g_5)+m_{\chi_i}(g_1g_3g_5+g_2g_4g_6)
\Big]\,C_1\\\non&+&3\Big[m_{\chi_i}(g_1g_3g_5+g_2g_4g_6)+
m_{\chi_j}(g_1g_3g_6+g_2g_4g_5)\Big]\,C_2\Big)\\
{\cal F}_3 &=& 8\Big(m_{\chi_i}(g_1g_3g_5 - g_2g_4g_6)\,C_0 + 
\Big[m_{\chi_k}(g_2g_3g_6-g_1g_4g_5)+
m_{\chi_i}(g_1g_3g_5-g_2g_4g_6)\Big]\,C_1\\\non&+
&\Big[m_{\chi_i}(g_1g_3g_5-g_2g_4g_6)-
m_{\chi_j}(g_2g_4g_5-g_1g_3g_6)\Big]\,C_2\Big)
\end{eqnarray}
The arguments of $B_i$ and $C_i$ functions are
$B_i(m^2_V,m^2_{\chi_i},m^2_{\chi_j})$, 
$C_i(m^2_W,m^2_V,\mhp^2,m^2_{\chi_i},m^2_{\chi_k},m^2_{\chi_j})$.
The couplings are given in the following table:
\begin{center}
\begin{tabular}{|ccc|c|c|c|c|c|c|}\hline
$\chi_i$&$\chi_j$&$\chi_k$&$g_1$&$g_2$&$g_3$&$g_4$&$g_5$&$g_6$ \\  \hline
$\tilde\chi^0$ & $\tilde\chi^0$&$\tilde\chi^+$
&$2\, \cO_{ij}^{\prime\prime{L}}/c_W$
&$-2\, \cO_{ij}^{\prime\prime{R}}/c_W$
&$-\, \cO_{kj}^{R*}$
&$-\, \cO_{kj}^{L*}$
&$-\,c_\beta\,\cO_{ki}^{\prime{L}}$
&$-\,s_\beta\,\cO_{ki}^{\prime{R}}$  \\
$\tilde\chi^+$&$\tilde\chi^-$&$\tilde\chi^0$
&$\, \cO_{ij}^{\prime{L}}/c_W$
&$\, \cO_{ij}^{\prime{R}}/c_W$
&$-\, \cO_{kj}^{R*}$
&$-\, \cO_{kj}^{L*}$
&$-\,c_\beta\,\cO_{ki}^{\prime{L}}$
&$-\,s_\beta\,\cO_{ki}^{\prime{R}}$  \\  
\hline
\end{tabular}
\end{center}
Diagram Fig.~\ref{hwz}.2:
\begin{eqnarray}
{\cal{F}}_1&=& 4\,N_C\, g_1^Vg_2^Hg_3^W\,C_{00}\\
{\cal{F}}_2&=& N_C\,g_1^Vg_2^Hg_3^W\,\Big(C_0+2(C_1 + 2C_{12} + C_2)\Big)
\end{eqnarray}

Where $N_C=3$ for scalar quarks and $N_C=1$ for scalar leptons, 
the arguments of $C_i$ and $C_{ij}$ functions are as
 $C_{i,ij}(m^2_W,m^2_V,m^2_{S_i},m^2_{S_k},m^2_{S_j})$.  
The couplings are given in the following table:
\begin{center}
\begin{tabular}{|ccc|c|c|c|c|}\hline
$S_i$&$S_j$&$S_k$&$g_1^{\gamma}$&$g_{1}^Z$&$g_2^{H}$&$g_3^{W}$\\ \hline 
$\snu_l$&$\snu_l$&$\tilde{l}_L$&$-g_{\gamma\snu\snu}$&$-g_{Z\snu\snu}$&$
g_{H\snu\tilde{l}}$&$g_{W\snu\tilde{l}}$  \\
$\tilde{l}_L$&$\tilde{l}_L$&$\snu_l$&$g_{\gamma\tilde{l}\tilde{l}}$&$
g_{Z\tilde{l}\tilde{l}}$&$g_{H\tilde{l}\snu}$&$g_{W\tilde{l}\snu}$ \\
$\tilde{u}_L$&$\tilde{u}_L$&$\tilde{d}_L$& $-g_{\gamma\tilde{u}\tilde{u}}$&$-
g_{Z\tilde{u}\tilde{u}}$&$g_{H\tilde{u}\tilde{d}}$&$g_{W\tilde{u}\tilde{d}}$\\
$\tilde{d}_L$&$\tilde{d}_L$&$\tilde{u}_L$&$g_{\gamma\tilde{d}\tilde{d}}$&$
g_{Z\tilde{d}\tilde{d}}$&$g_{H\tilde{d}\tilde{u}}$&$g_{W\tilde{d}\tilde{u}}$ \\
$\tilde{t}_i$&$\tilde{t}_j$&$\tilde{b}_k$&
$-\gpul{R}^t_{1i}{R}^t_{1j}-\gpur{R}^t_{2i}{R}^t_{2j}$&
$-\gzul{R}^t_{1i}{R}^t_{1j}-\gzur{R}^t_{2i}{R}^t_{2j}$&
$g_{H\tilde{b}\tilde{t}}$&$-{R}^t_{1j}{R}^b_{1k}$ \\
$\tilde{b}_i$&$\tilde{b}_i$&$\tilde{t}_k$&
$\gpdl{R}^b_{1i}{R}^b_{1j}+\gpdr{R}^b_{2i}{R}^b_{2j}$&
$\gzdl{R}^b_{1i}{R}^b_{1j}+\gzdr{R}^b_{2i}{R}^b_{2j}$&
$g_{H\tilde{b}\tilde{t}}$&${R}^t_{1k}{R}^b_{1j}$ \\ \hline
\end{tabular}
\end{center}
where 
$R^{t,b}_{11}=R^{t,b}_{22}=\cat$ and $-R^{t,b}_{12}=R^{t,b}_{21}=\sat$.\\

Diagram Fig.~\ref{hwz}.8:
\begin{eqnarray}
{\cal{F}}_1&=&-N_C g_1g_2 B_{0}(\mhp^2, m_{S_i}^2, m_{S_j}^2)
\end{eqnarray}

The couplings are given in the following table:

\begin{center}
\begin{tabular}{|cc|c|c|}\hline
$S_i$&$S_j$&$g_1$&$g_2$\\  \hline
$\tilde\nu_l$&$\tilde{l}_L$
&$g_{H^\pm\snu_l\tilde{l}_L}$
&$g_{\snu\tilde{l}_L W^\pm V}$\\
$\tilde{t}_L$&$\tilde{b}_L$
&$g_{H^\pm\tilde{t}_L\tilde{b}_L}$
&$g_{\tilde{t}_L\tilde{b}_L W^\pm V}$\\
$\tilde{t}_1$&$\tilde{b}_1$
&$g_{H^\pm\tilde{t}_1\tilde{b}_1}$
&$g_{\tilde{t}_1\tilde{b}_1 W^\pm V}$\\
$\tilde{t}_1$&$\tilde{b}_2$
&$g_{H^\pm\tilde{t}_1\tilde{b}_2}$
&$g_{\tilde{t}_1\tilde{b}_2 W^\pm V}$\\
$\tilde{t}_2$&$\tilde{b}_1$
&$g_{H^\pm\tilde{t}_2\tilde{b}_1}$
&$g_{\tilde{t}_2\tilde{b}_1 W^\pm V}$\\
$\tilde{t}_2$&$\tilde{b}_2$
&$g_{H^\pm\tilde{t}_2\tilde{b}_2}$
&$g_{\tilde{t}_2\tilde{b}_2 W^\pm V}$\\
\hline
\end{tabular}
\end{center}

Diagram Fig.~\ref{hwz}.12:
\begin{eqnarray}
{\cal{F}}_2&=&\frac{2\,N_C\,g_3}{m_{W}^2-\mhp^2}\Big(-m_{F_i}
(g_1g_4+g_2g_5)B_0 + \Big[(g_1g_4+g_2g_5)m_{F_i}+(g_2g_4+g_1g_5)m_{F_j}\Big]B_1
\Big)
\end{eqnarray}
The arguments for the Passarino-Veltman functions 
$B_{i}$ are $B_i(m_V^2,m^2_{F_i},m^2_{F_j})$.

The couplings are given in the following table:
\begin{center}
\begin{tabular}{|ccc|c|c|c|c|c|}\hline
$F_i$&$F_j$&$S_k$&$g_1$&$g_2$&$g_3$&$g_4$&$g_5$ \\  \hline
$t$&$b$&$H^\pm$
&$0$
&$g_{W^\pm tb}^L$
&$g_{H^{\pm}W^\pm V}$
&$g_{H^\pm tb}^L$
&$g_{H^\mp tb}^R$\\
$\tilde\chi^0$&$\tilde\chi^+$&$H^\pm$
&$\cO^{L*}_{ji}$
&$\cO^{R*}_{ji}$
&$g_{H^{\pm}W^\pm V}$
&$-c_{\beta}\cO^{\prime L}_{ji}$
&$-s_{\beta}\cO^{\prime R}_{ji}$
\\
\hline
\end{tabular}
\end{center}
Diagram Fig.~\ref{hwz}.13:
\begin{eqnarray}
{\cal{F}}_2=\frac{N_C\,g_1\,g_2\,g_3}{m_{W}^2 - \mhp^2}
 \Big(B_0(m_{W}^2, m_{S_j}^2, m_{S_i}^2) + 
2 B_1(m_{W}^2,m_{S_j}^2, m_{S_i}^2 )\Big)
\end{eqnarray}
The couplings are given in the following table:
\begin{center}
\begin{tabular}{|ccc|c|c|c|}\hline
$S_i$&$S_j$&$S_k$&$g_1$&$g_2$&$g_3$\\  \hline
$\tilde\nu_l$&$\tilde{l}_L$&$H^\pm$
&$g_{H^\pm H^\mp V}$
&$g_{H^\pm\snu_l\tilde{l}_L}$
&$g_{\snu_l\tilde{l}_L W^\pm }$\\
$\tilde{u}_L$&$\tilde{d}_L$&$H^\pm$
&$g_{H^\pm H^\mp V}$
&$g_{H^\pm\tilde{u}_L\tilde{d}_L}$
&$g_{\tilde{u}_L\tilde{d}_L W^\pm }$\\
$\tilde{t}_1$&$\tilde{b}_1$&$H^\pm$
&$g_{H^\pm H^\mp V}$
&$g_{H^\pm\tilde{t}_1\tilde{b}_1}$
&$g_{\tilde{t}_1\tilde{b}_1 W^\pm }$\\
$\tilde{t}_1$&$\tilde{b}_2$&$H^\pm$
&$g_{H^\pm H^\mp V}$
&$g_{H^\pm\tilde{t}_1\tilde{b}_2}$
&$g_{\tilde{t}_1\tilde{b}_2 W^\pm }$\\
$\tilde{t}_2$&$\tilde{b}_1$&$H^\pm$
&$g_{H^\pm H^\mp V}$
&$g_{H^\pm\tilde{t}_2\tilde{b}_1}$
&$g_{\tilde{t}_2\tilde{b}_1 W^\pm }$\\
$\tilde{t}_2$&$\tilde{b}_2$&$H^\pm$
&$g_{H^\pm H^\mp V}$
&$g_{H^\pm\tilde{t}_2\tilde{b}_2}$
&$g_{\tilde{t}_2\tilde{b}_2 W^\pm }$\\
\hline
\end{tabular}
\end{center}

Diagram Fig.~\ref{hwz}.15:
\begin{eqnarray}
{\cal{F}}_1&=&\frac{2\,N_C\,g_5}{\mhp^2 - m_{W}^2}\Big(
(g_2g_3 + g_1g_4)\,A_{0}(m_{F_j}^2) +
 m_{F_i}\Big[(g_2g_3 + g_1g_4)m_{F_i} + (g_1g_3+ g_2g_4)m_{F_j}\Big]B_0 
\\\non&+&(g_2g_3 + g_1g_4)g_5\mhp^2 B_1\Big)
\end{eqnarray}
The arguments of $B_{i}$ functions are 
$B_i(\mhp^2,m^2_{F_i},m^2_{F_j})$. 
The couplings are given in the following table:
\begin{center}
\begin{tabular}{|ccc|c|c|c|c|c|}\hline
$F_i$&$F_j$&$S_k$&$g_1$&$g_2$&$g_3$&$g_4$&$g_5$ \\  \hline
$t$&$b$&$G^\pm$
&$g_{G^\pm tb}^L$
&$g_{G^\pm tb}^R$
&$g_{G\pm W^\pm V}$
&$g_{H^\pm tb}^L$
&$g_{H^\mp tb}^R$\\
$\tilde\chi^0$&$\tilde\chi^+$&$G^\pm$
&$-c_{\beta}\cO^{\prime L}_{ji}$
&$-s_{\beta}\cO^{\prime R}_{ji}$
& $g_{G^\pm W^\mp V}$
&$c_{\beta}\cO^{\prime R*}_{ji}$
&$-s_{\beta}\cO^{\prime L*}_{ji}$ \\
\hline
\end{tabular}
\end{center}
Diagram Fig.~\ref{hwz}.16:
\begin{eqnarray}
{\cal{F}}_1=\frac{N_Cg_1g_2g_3}{\mhp^2-m_{S_k}^2}B_{0}
(\mhp^2,m_{S_j}^2,m_{S_i}^2)
\end{eqnarray}
The couplings are given in the following table:
\begin{center}
\begin{tabular}{|ccc|c|c|c|}\hline
$S_i$&$S_j$&$S_k$&$g_1$&$g_2$&$g_3$\\  \hline
$\tilde\nu_l$&$\tilde{l}_L$&$G^\pm$
&$g_{H^\pm\snu_l\tilde{l}_L}$
&$g_{G^\pm W^\pm V}$
&$g_{G^\pm\snu_l\tilde{l}_L}$\\
$\tilde{u}_L$&$\tilde{d}_L$&$G^\pm$
&$g_{H^\pm\tilde{u}_L\tilde{d}_L}$
&$g_{G^\pm W^\pm V}$
&$g_{G^\pm\tilde{u}_L\tilde{d}_L}$\\
$\tilde{t}_1$&$\tilde{b}_1$&$G^\pm$
&$g_{H^\pm\tilde{t}_1\tilde{b}_1}$
&$g_{G^\pm W^\pm V}$
&$g_{G^\pm\tilde{t}_1\tilde{b}_1}$\\
$\tilde{t}_1$&$\tilde{b}_2$&$G^\pm$
&$g_{H^\pm\tilde{t}_1\tilde{b}_2}$
&$g_{G^\pm W^\pm V}$
&$g_{G^\pm\tilde{t}_1\tilde{b}_2}$\\
$\tilde{t}_2$&$\tilde{b}_1$&$G^\pm$
&$g_{H^\pm\tilde{t}_2\tilde{b}_1}$
&$g_{G^\pm W^\pm V}$
&$g_{G^\pm\tilde{t}_2\tilde{b}_1}$\\
$\tilde{t}_2$&$\tilde{b}_2$&$G^\pm$
&$g_{H^\pm\tilde{t}_2\tilde{b}_2}$
&$g_{G^\pm W^\pm V}$
&$g_{G^\pm\tilde{t}_2\tilde{b}_2}$\\
\hline
\end{tabular}
\end{center}

Diagram Fig.~\ref{hwz}.18:
\begin{eqnarray}
{\cal{F}}_2&=&\frac{N_C\,g_3}{\mhp^2-m_{W}^2}\Big(B_0(g_2g_4 + g_1g_5)
m_{F_i}+\Big[(g_2g_4 + g_1g_5)m_{F_i} + (g_1g_4 + g_2g_5)m_{F_j}\Big]B_1\Big)\\
{\cal{F}}_1&=&\Big(\mhp^2-m_{V}^2\Big){\cal F}_2
\end{eqnarray}

The arguments of $B_{i}$ functions are as
$B_i(\mhp^2,m^2_{F_i},m^2_{F_j})$.
The couplings are given in the following table:

\begin{center}
\begin{tabular}{|ccc|c|c|c|c|c|}\hline
$F_i$&$F_j$&$V_k$&$g_1$&$g_2$&$g_3$&$g_4$&$g_5$ \\  \hline
$t$&$b$&$W^\pm$
&$g_{W^\pm tb}^L$
&$0$
&$g_{W^\pm W^\pm V}$
&$g_{H^\pm tb}^L$
&$g_{H^\pm tb}^R$\\
$\tilde\chi^0$&$\tilde\chi^+$&$W^\pm$
&$g_{W^\pm\tilde\chi^\pm_j\tilde\chi^0_i}^L$
&$g_{W^\pm\tilde\chi^\pm_j\tilde\chi^0_i}^R$
&$g_{W^\pm W^\pm V}$
&$-c_{\beta}\cO^{\prime L}_{ji}$
&$-s_{\beta}\cO^{\prime R}_{ji}$\\
\hline
\end{tabular}
\end{center}
Diagram Fig.~\ref{hwz}.19:
\begin{eqnarray}
{\cal{F}}_1&=&(m^2_V - m^2_W){\cal{F}}_2\\
{\cal{F}}_2 &=&\frac{g_1g_2g_3}{\mhp^2 - m_{V_k}^2}\Big(B_0+2 B_1\Big)
\end{eqnarray}
The arguments of $B_i$  functions are as
$B_{i}(\mhp^2,m_{S_i}^2,m_{S_j}^2)$.

The couplings are given in the following table:
\begin{center}
\begin{tabular}{|ccc|c|c|c|}\hline
$S_i$&$S_j$&$V_k$&$g_1$&$g_2$&$g_3$\\  \hline
$\tilde\nu_l$&$\tilde{l}_L$&$W^\pm$
&$g_{H^\pm\snu_l\tilde{l}_L}$
&$g_{W^\pm W^\pm V}$
&$g_{W^\pm\snu_l\tilde{l}_L}$\\
$\tilde{u}_L$&$\tilde{d}_L$&$W^\pm$
&$g_{H^\pm\tilde{u}_L\tilde{d}_L}$
&$g_{W^\pm W^\pm V}$
&$g_{W^\pm\tilde{u}_L\tilde{d}_L}$\\
$\tilde{t}_1$&$\tilde{b}_1$&$W^\pm$
&$g_{H^\pm\tilde{t}_1\tilde{b}_1}$
&$g_{W^\pm W^\pm V}$
&$g_{W^\pm\tilde{t}_1\tilde{b}_1}$\\
$\tilde{t}_1$&$\tilde{b}_2$&$W^\pm$
&$g_{H^\pm\tilde{t}_1\tilde{b}_2}$
&$g_{W^\pm W^\pm V}$
&$g_{W^\pm\tilde{t}_1\tilde{b}_2}$\\
$\tilde{t}_2$&$\tilde{b}_1$&$W^\pm$
&$g_{H^\pm\tilde{t}_2\tilde{b}_1}$
&$g_{W^\pm W^\pm V}$
&$g_{W^\pm\tilde{t}_2\tilde{b}_1}$\\
$\tilde{t}_2$&$\tilde{b}_2$&$W^\pm$
&$g_{H^\pm\tilde{t}_2\tilde{b}_2}$
&$g_{W^\pm W^\pm V}$
&$g_{W^\pm\tilde{t}_2\tilde{b}_2}$\\
\hline
\end{tabular}
\end{center}
{\bf{Counter-term and self energies diagrams:}}\\
As explained in section.II-B, the counter-term $\Delta$ is fixed by 
real part of $W^\pm H^\pm$ self energy mixing eq.\ref{Del}.
Hereafter we list the fermion $\Sigma_{HW}^f$ and sfermion 
$\Sigma_{HW}^S$ contribution to 
$\Sigma_{HW}=\Sigma_{HW}^f + \Sigma_{HW}^S$.

Diagram Fig.~\ref{hwmix}.1a:
\begin{eqnarray}
\Sigma_{HW}^{f}(k^2) &=&
\frac{N_c}{8\pi^2}
\{\Big(m_{b}(g_{W}^Lg_{H}^R + g_{W}^Rg_{H}^L) + m_{t}(g_{W}^Lg_{H}^L + 
g_{W}^Rg_{H}^R)\Big)B_1\non\\&+& 
m_{b}(g_{W}^Lg_{H}^R + g_{W}^Rg_{H}^L)B_0) \}
\end{eqnarray}
Where $k$ is the momentum of the external particle, $g_{W,H}=g_{Wtb,Htb}$, 
the arguments of $B_i$ functions are as $B_i(k^2,m^2_{b},m^2_{t})$.
\begin{center}
\begin{tabular}{|cc|c|c|c|c|} \hline
$F$&$F$&$g_{H}^L$&$g_{H}^R$&$g_{W}^L$&$g_{W}^R$ \\ \hline
$t$&$b$
&$g_{Htb}^L$&$g_{Htb}^R$&$g_{Wtb}^L$&$g_{Wtb}^R$\\
$\tilde\chi^0$&$\tilde\chi^\pm$
&$g_{H\tilde\chi^0\tilde\chi^\pm}^L$&$g_{H\tilde\chi^0\tilde\chi^\pm}^R$
&$g_{W\tilde\chi^0\tilde\chi^\pm}^L$&$g_{W\tilde\chi^0\tilde\chi^\pm}^R$
\\
\hline
\end{tabular}
\end{center}
Diagram Fig.~\ref{hwmix}.1b:
\begin{eqnarray}
\Sigma_{HW}^{S}(k^2)&=&
\frac{N_c}{16\pi^2}g_{HS_iS_j}g_{WS_iS_j}\Big(B_0 + 2B_1\Big)
\end{eqnarray}
The arguments of $B_i$ functions are as $B_i(k^2,m^2_{S_i},m^2_{S_j})$.
\begin{center}
\begin{tabular}{|cc|c|c|} \hline
$S_i$&$S_j$&$g_{HS_iS_j}$&$g_{WS_iS_j}$ \\ \hline
$\tilde{\nu}_l$&$\tilde{l}_L$
&$g_{H\tilde{\nu}_l\tilde{l}_L}$
&$g_{W\tilde{\nu}_l\tilde{l}_L}$\\
$\tilde{u}_L$&$\tilde{d}_L$
&$g_{H\tilde{u}_L\tilde{d}_L}$
&$g_{W\tilde{u}_L\tilde{d}_L}$\\
$\tilde{t}_1$&$\tilde{b}_1$
&$g_{H\tilde{t}_1\tilde{b}_1}$
&$g_{W\tilde{t}_1\tilde{b}_1}$\\
$\tilde{t}_1$&$\tilde{b}_2$
&$g_{H\tilde{t}_1\tilde{b}_2}$
&$g_{W\tilde{t}_1\tilde{b}_2}$\\
$\tilde{t}_2$&$\tilde{b}_1$
&$g_{H\tilde{t}_2\tilde{b}_1}$
&$g_{W\tilde{t}_2\tilde{b}_1}$\\
$\tilde{t}_2$&$\tilde{b}_2$
&$g_{H\tilde{t}_2\tilde{b}_2}$
&$g_{W\tilde{t}_2\tilde{b}_2}$\\
\hline
\end{tabular}
\end{center}
Diagram Fig.~\ref{hwmix}.1c:
\begin{eqnarray}
\Sigma_{HG}^f(k^2)&=&
\frac{N_c}{8\pi^2}\Big(
(g_{H}^Lg_{G}^R + g_{H}^Rg_{G}^L)\,A_{0}(m_{F_j}^2) +
 m_{F_i}((g_{H}^Lg_{G}^R + g_{H}^Rg_{G}^L)m_{F_i} \non\\&+& 
(g_{H}^Rg_{G}^R + g_{H}^Lg_{G}^L)m_{F_j})B_0 +
(g_{H}^Lg_{G}^R + g_{H}^Rg_{G}^L)k^2 B_1\Big)
\end{eqnarray}
The arguments of $B_i$ functions are as $B(k^2,m^2_{F_i},m^2_{F_j})$.
\begin{center}
\begin{tabular}{|cc|c|c|c|c|} \hline
$F_i$&$F_j$&$g_{H}^L$&$g_{H}^R$&$g_{G}^L$&$g_{G}^R$ \\ \hline
$t$&$b$
&$g_{Htb}^L$&$g_{Htb}^R$&$g_{Gtb}^L$&$g_{Gtb}^R$\\
$\tilde\chi^0$&$\tilde\chi^\pm$
&$g_{H\tilde\chi^0\tilde\chi^\pm}^L$&$g_{H\tilde\chi^0\tilde\chi^\pm}^R$
&$g_{G\tilde\chi^0\tilde\chi^\pm}^L$&$g_{G\tilde\chi^0\tilde\chi^\pm}^R$
\\
\hline
\end{tabular}
\end{center}
Diagram Fig.~\ref{hwmix}.1d:
\begin{eqnarray}
\Sigma_{HG}^S=\frac{-N_c}{16\pi^2}g_{HS_iSj}g_{GS_iS_j}
B_{0}(k^2,m_{S_j}^2,m_{S_i}^2)
\end{eqnarray}

\begin{center}
\begin{tabular}{|cc|c|c|} \hline
$S_i$&$S_j$&$g_{HS_iS_j}$&$g_{GS_iS_j}$ \\ \hline
$\tilde{\nu}_l$&$\tilde{l}_L$
&$g_{H\tilde{\nu}_l\tilde{l}_L}$&$g_{G\tilde{\nu}_l\tilde{l}_L}$\\
$\tilde{u}_L$&$\tilde{d}_L$
&$g_{H\tilde{u}_L\tilde{d}_L}$
&$g_{G\tilde{u}_L\tilde{d}_L}$\\
$\tilde{t}_1$&$\tilde{b}_1$
&$g_{H\tilde{t}_1\tilde{b}_1}$
&$g_{G\tilde{t}_1\tilde{b}_1}$\\
$\tilde{t}_1$&$\tilde{b}_2$
&$g_{H\tilde{t}_1\tilde{b}_2}$
&$g_{G\tilde{t}_1\tilde{b}_2}$\\
$\tilde{t}_2$&$\tilde{b}_1$
&$g_{H\tilde{t}_2\tilde{b}_1}$
&$g_{G\tilde{t}_2\tilde{b}_1}$\\
$\tilde{t}_2$&$\tilde{b}_2$
&$g_{H\tilde{t}_2\tilde{b}_2}$
&$g_{G\tilde{t}_2\tilde{b}_2}$\\
\hline
\end{tabular}
\end{center}
The amplitudes of the counter-terms Fig.~2.2a $\to$ d, are given by:\\ 
Diagram Fig.~\ref{hwmix}.2a:
\begin{eqnarray}
{\cal F}_1^{\gamma}= e\Delta \qquad
{\cal F}_1^Z= e\frac{\sw}{\cw}\Delta
\end{eqnarray}
Diagram Fig.~\ref{hwmix}.2b:
\begin{eqnarray}
{\cal F}_2 = -g_{HHV} \Delta
\end{eqnarray}
Diagram Fig.~\ref{hwmix}.2c:
\begin{eqnarray}
{\cal F}_1= \frac{g_{HGV}}{\mw}\frac{\mhp^4}{\mhp^2-\mw^2} \Delta
\end{eqnarray}
Diagram Fig.~\ref{hwmix}.2d:
\begin{eqnarray}
{\cal F}_1 & =& \frac{g_{WWV}}{\mhp^2-\mw^2} (\mw^2-m^2_V)
\Delta\nonumber\\
{\cal F}_2 & =& -\frac{g_{WWV}}{\mhp^2-\mw^2}\Delta
\end{eqnarray}
where $\Delta$ is fixed by eq.\ref{Del} and the couplings
$g_{HHV}$, $g_{HGV}$ and $g_{WWV}$ have been defined above.

\end{document}